\documentclass[10pt,a4paper]{article}
\usepackage{jheppub_kim}
\usepackage{pdflscape}
\usepackage{amsmath}
\usepackage{amssymb}
\usepackage{dcolumn}
\usepackage{bm}
\usepackage{color}
\usepackage{epsfig}
\usepackage{amsfonts}
\usepackage{graphicx}
\usepackage{subfigure}
\usepackage{dcolumn}
\UseRawInputEncoding

\begin{document}
\title{Energy-momentum squared symmetric Teleparallel gravity: $f(Q,T_{\mu\nu}T^{\mu\nu})$ gravity}

\author[a]{Prabir Rudra}

\affiliation[a] {Department of Mathematics, Asutosh College,
Kolkata-700 026, India}

\emailAdd{prudra.math@gmail.com, prabir.rudra@asutoshcollege.in}

\abstract{In this work we propose the $f(Q,T_{\mu\nu}T^{\mu\nu})$
gravity as a further extension of the $f(Q)$ and $f(Q,T)$ gravity
theories, where $Q$ is the non-metricity and $T_{\mu\nu}$ is the
energy-momentum tensor. The action involves an arbitrary function
of the non-metricity $Q$ and $\textbf{T}^2=T_{\mu\nu}T^{\mu\nu}$
in the gravity Lagrangian. The field equations for the theory are
derived in the metric-affine formalism. The theory involves a
non-minimal coupling between the geometric and the matter sectors,
and hence the covariant divergence of the energy momentum tensor
is non-zero, thus implying the non-conservation of the same. The
vacuum solutions of the theory are investigated and it is found
that the theory perfectly admits a de-Sitter-like evolution of the
universe. The cosmological equations are derived and it is found
that there are two correction terms arising as modification of the
gravity. Two specific toy models of the form $Q+\eta
\left(\textbf{T}^2\right)^{n}$ and $f_{0}Q^{m}
\left(\textbf{T}^2\right)^{n}$ are explored to gain further
insights into the dynamics of the theory. It is seen that the
field equations of both the models have terms similar to those
arising from the quantum gravity effects and are thus responsible
for the avoidance of the singularity. One striking feature of the
model is that the non-linear correction terms dominate in the
early universe and gradually fade away at later times giving
standard FLRW universe. Solutions for the FLRW equations are found
wherever possible and the evolution of the scale factor and the
matter energy density is plotted. Other cosmological parameters
like the equation of state, deceleration parameters and Hubble
functions are also studied. Finally the energy conditions are
explored in the background of the theory. Using these conditions
and some observational data the parameter spaces of the models are
considerably constrained. $f(Q,T_{\mu\nu}T^{\mu\nu})$ is a theory
that can perfectly explain the cosmological dynamics of both the
early and the late universe without resorting to any dark energy.}

\keywords{non-metricity, energy-momentum, symmetric teleparallel
gravity; cosmology, energy conditions.}

\maketitle

\section{Introduction}
Currently the most comprehensive theory of gravity available to us
is the theory of General Relativity (GR) \cite{gr} proposed by
Albert Einstein in 1916. After a century of extensive research GR
has been able to survive quite well with many of its predictions
coming true. Some of them being the perihelion precession of
mercury, deflection of light by sun \cite{eddo}, gravitational
redshift \cite{redshift1}, detection of gravitational waves
\cite{gw1} from the mergers of black holes and neutron stars,
imaging a black hole shadow in the M87 galaxy by the Event horizon
telescope \cite{ehtc}, etc. Extensive reviews in experimental and
observational tests of GR can be found in \cite{tests1, tests2}.
Nevertheless some observations did not go in favour of GR, quite
expectedly. The most important one being the discovery of the late
cosmic acceleration \cite{acc1, acc2} at the turn of the last
century. With this observation to light, GR became incompatible at
cosmological distances. One more important observation was that GR
was not able to explain the gravitational interaction at the
quantum level. Moreover standard cosmology is plagued by problems
like the singularity problem, cosmological constant problem,
cosmic coincidence problem, etc. So it is clear that we are far
from the theory that we need to comprehensively explain our
observations. This provided motivation for extensive theoretical
research in cosmology over the past century. With the development
of quantum mechanics throughout the last century, scientists
including Einstein took up the challenge of developing a proper
theory of quantum gravity. As a result theories like string
theory, loop quantum gravity theory, etc. were proposed, but till
date they are far from being comprehensive. To resolve the
incompatibility of GR at cosmological scales mainstream research
has travelled in two different ways. The first path is the theory
of dark energy (DE) where the matter content of the universe is
modelled by an exotic fluid with negative pressure driving the
acceleration. Extensive reviews on DE can be found in \cite{de1}.
The alternative path is the theory of modified gravity, where the
gravitational framework of GR has been modified to incorporate the
cosmic acceleration. See \cite{mod1, mod2, mod3} for extensive
reviews on modified gravity. In this work we will concentrate on
this alternative path of modified gravity theory.

The field equations of GR are basically formulated from the
Einstein-Hilbert (EH) action
$\mathcal{S}=\int\left(R+\mathcal{L}_{m}\right)\sqrt{-g}d^{4}x$,
where $R$ is the Ricci scalar, $g$ is the determinant of the
metric tensor $(g_{\mu\nu})$ that represents the gravitational
field and $\mathcal{L}_{m}$ is the matter Lagrangian. The most
fundamental form of modification is imposed by replacing the
gravitational Lagrangian $R$ of EH action by an arbitrary function
$f(R)$, thus considering a generalized action. This gave rise to
$f(R)$ theories \cite{fr1}. Substantive development in $f(R)$
gravity can be found in \cite{fr2, fr3, fr4, fr5, fr6, fr7}.
Comprehensive reviews on this theory can be found in
Refs.\cite{frrev1, frrev2}. In similar fashion further extensions
to the EH action was brought about by replacing the arbitrary
function $f(R)$ by $f(R,\mathcal{L}_{m})$ \cite{frlm1, frlm2,
frlm3, frlm4}. In ref.\cite{harko1} the authors modified the EH
action by including an arbitrary function $f(R,T)$, where $T$ is
the trace of the energy-momentum tensor (EMT). Here a coupling
between the matter sector and the gravity sector was considered
via the function $f(R,T)$. It was seen that the covariant
divergence of the EMT is non-zero for this theory, which means
non-conservation of EMT leading to non-geodesic motion of the
massive test particles. The reason for this being the coupling
effects between matter and geometry which induces extra
acceleration on the particles. Further developments in $f(R,T)$
gravity can be found in \cite{frt1, frt2, frt3, frt4, frt5, frt6,
frt7}. Katirci and Kavuk in \cite{emsgorg1} proposed $f(R,
\textbf{T}^{2})$ theory, where
$\textbf{T}^{2}=T_{\mu\nu}T^{\mu\nu}$ and $T_{\mu\nu}$ is the EMT.
Roshan and Shojai in \cite{emsgorg} further developed the theory
by investigating the properties of a specific form $R+\eta
\textbf{T}^2$ which was termed as energy-momentum squared gravity
(EMSG) in the literature. From the field equations of EMSG it was
seen that the equations have a flavour of the quantum geometry
effects of loop quantum gravity \cite{lqg1}, and allowed a maximum
energy density $\rho_{max}$ and a minimum length $a_{min}$ in the
early universe. As a result, in a homogeneous and isotropic
spacetime this theory admits a cosmological bounce and avoids the
existence of the early time singularity. Cosmological models in
EMSG were studied by the authors in \cite{board1, akarsu2}. An
extensive dynamical system analysis in EMSG was presented by the
authors of \cite{rudra1}. Observational constraints on EMSG using
cosmic chronometers and supernova observations can be found in
\cite{rudra2}. Thermodynamics of the apparent horizon in the
generalized energy-momentum squared cosmology was studied in
\cite{rudra3}. Other substantial developments in EMSG can be found
in \cite{akarsu3, akarsu1, moraes1, nari1, keskin1, akarsu4}.

All the above extensions of GR have one feature common between
them, which is the underlying Riemannian geometry \cite{riemanng}
(formulated in the Riemann metrical space) lying at the heart of
all these classical theories including GR. With the
incompatibility of these theories at some scales, naturally a
thought arises that if the underlying geometry can be replaced by
a far more general geometric structure, then we may be able to
remove some of the inconsistencies that has plagued these
classical theories over the years. Such a novel attempt was
undertaken by Weyl \cite{weyl}, where the main objective was the
geometrical unification of gravity and electromagnetism. We know
that in Riemannian geometry, the Levi-Civita connection is
compatible with the metric and is the basic tool for length
comparison between vectors. In Weyl's theory there is a complete
different mechanism where two connections are used, one bearing
the information of the length of a vector and the other
responsible for the direction of a vector during parallel
transport. Physically the length connection is identified with the
electromagnetic potential. The most striking feature of the theory
being the non-zero covariant divergence of the metric tensor and
this property induces a new geometrical quantity known as the
\textit{non-metricity}. Weyl's geometry is a mathematical
masterpiece with corresponding rich physical structure.

Since gravity is identified as the manifestation of the
geometrical properties of spacetime, search for other developments
in geometry continued. Based on the works \cite{cartan1, cartan2,
cartan3, cartan4} Cartan proposed an extension of GR known as the
Einstein-Cartan theory. A review on these theories may be found in
\cite{ecrev}. The striking feature of Cartan's geometry was the
introduction of the \textit{torsion field}, which is interpreted
as the spin density from the physical point of view \cite{ecrev}.
This torsion field may be introduced in Weyl's geometric structure
to give a natural extension of Weyl-Cartan geometry, which can be
an interesting mathematical structure with associated physical
implications \cite{wc1, wc2, wc3, wc4}. An extensive review on
Riemann-Cartan and Weyl-Cartan geometries can be found in
\cite{ecwcrev}. Another elegant geometrical formalism was given by
Weitzenbock known as the Weitzenbock spaces \cite{weit}, where the
manifold is equipped with the properties
$R^{\alpha}_{~\beta\gamma\delta}=0$,
$T^{\alpha}_{~\beta\gamma}\neq 0$,
$\nabla_{\mu}g_{\alpha\beta}=0$, where
$R^{\alpha}_{~\beta\gamma\delta}$ is the curvature tensor,
$T^{\alpha}_{~\beta\gamma}$ is the torsion tensor and
$g_{\alpha\beta}$ is the metric tensor of the associated manifold.
If the torsion tensor vanishes the Weitzenbock manifold reduces to
the Euclidean manifold. Since the Weitzenbock space is
curvature-less, the geometry derived from such spaces possesses an
important property of absolute parallelism or teleparallelism.
This property of Weitzenbock spaces attracted Einstein, who
applied this theory in physics by proposing a unified teleparallel
theory of gravity and electromagnetism \cite{tegr1}. In
teleparallel gravity the basic idea is to replace the metric
$g_{\alpha\beta}$ by the tetrad vectors $e^{j}_{\mu}$, which
generates the torsion that is responsible for the gravitational
effects. Since in this theory the concept of torsion exactly
replaces the concept of curvature, it is termed as the
teleparallel equivalent of general relativity (TEGR) \cite{tegr2,
tegr3, tegr4}. One important thing to note in TEGR is that the
spacetime in these theories are flat, due to the absence of
curvature. Just like GR was extended to $f(R)$ theories, similarly
TEGR was extended to $f(\tau)$ theories \cite{ft1, ft2}, where
$\tau$ is the scalar torsion. One significant advantage of
$f(\tau)$ theories over $f(R)$ theories lies in the fact that the
field equations for the former are of the second order in contrast
to the fourth order field equations of the latter. More
importantly the $f(\tau)$ theories have been successful in
explaining various astrophysical processes and specifically the
late cosmic acceleration without depending on any forms of exotic
matter or dark energy \cite{fta1, fta2, fta3, fta4, fta5, fta6,
fta7, fta8, fta9}. A recent review on Teleparallel gravity can be
found in \cite{revsb}. Further extensions of the teleparallel
framework was done via the Weyl-Cartan-Weitzenbock (WCW) theory
\cite{wcw1, wcw2}, where the extra condition of the exact
cancellation of curvature by the torsion is introduced in the
background of the Weyl-Cartan spacetime.

From the above discussions we have seen that there are two
alternative formulations of GR: the first one with $R\neq 0,
\tau=0$ (curvature formulation), and the second one with $R=0,
\tau\neq 0$ (teleparallel formulation). However in both these
formulations the non-metricity $Q$ vanishes. Geometrically $Q$
represents the variation of length of a vector in parallel
transport. Now in a third equivalent formalism of GR, a
non-vanishing non-metricity $Q$ was considered as the basic
geometrical variable responsible for all forms of gravitational
interactions. This theory was termed the \textit{symmetric
teleparallel gravity} (STG) \cite{st1}. Here the Einstein
pseudotensor plays the role of the energy-momentum density and in
the geometric representation it finally becomes a true tensor.
Further research saw the STG being extended to $f(Q)$ gravity
\cite{cgr1}, which is also known as the \textit{coincident general
relativity} and \textit{non-metric gravity}. Cosmology of $f(Q)$
gravity and its observational constraints was studied in
\cite{fq1, fq2}. Over the past decades there have been various
developments in the framework of STG \cite{stegr2, stegr3, stegr4,
stegr5, dsit}. In ref.\cite{dsit} the authors have proposed an
extension of the $f(Q)$ gravity based on the non-minimal coupling
between the non-metricity $Q$ and the matter Lagrangian
$\mathcal{L}_{m}$. Quite expectedly the non-minimal coupling
between geometry and matter sectors results in the
non-conservation of the energy-momentum tensor and appearance of
an extra force in the geodesic equation of motion. Xu et. al in
\cite{ftq1} developed another extension of the theory namely the
$f(Q,T)$ gravity where the gravity Lagrangian is basically an
arbitrary function of $Q$ and the trace of the energy-momentum
tensor $T$. The field equations were derived and the cosmological
evolution of the model was studied. It was seen that in all the
considered cases the theory supported an accelerated expansion of
the universe ending with a de-Sitter type evolution.

In this work we are motivated to explore another extension of the
symmetric teleparallel theory. Our basic aim is to extend $f(Q,T)$
gravity to $f(Q,\textbf{T}^{2})$ gravity, where
$\textbf{T}^{2}=T_{\mu\nu}T^{\mu\nu}$, following the extension of
$f(R,T)$ gravity to $f(R,\textbf{T}^2)$ gravity. The novel
features of $f(R,\textbf{T}^2)$ gravity, would be a direct
motivation to perform this analogical study. A second motivation
would be to develop the $f(R,\mathcal{L}_m)$ theory where matter
is non-minimally coupled to geometry. Now the matter Lagrangian
can be generated via various physical invariants. The most common
is the trace of the energy momentum tensor
$T=g^{\mu\nu}T_{\mu\nu}$ which led to $f(R,T)$ theory. The next
obvious development would be to incorporate some extensions of $T$
and check its evolution in cosmology. This is exactly what we have
done by extending to $T$ to $\mathbf{T}^2=T_{\mu\nu} T^{\mu\nu}$,
which is perhaps the simplest logical extension, whose evolution
is worth studying due to its non linear nature in the matter
sector. An important feature of the $f(R,\mathcal{L}_m)$ theory
and its derivatives is that they have connections with MOND
(modified Newtonian dynamics) \cite{mond1} which is an alternative
to the dark matter hypothesis explaining why the dynamics of
galaxies do not obey the currently understood laws of physics.
This is a huge motivation to study these classes of models.

We will call $f(Q,\textbf{T}^{2})$ gravity as the Energy-momentum
squared symmetric teleparallel gravity (EMSSTG). The gravitational
action will be constituted by an arbitrary function
$f(Q,\textbf{T}^{2})$ of $Q$ and $\textbf{T}^{2}$. Then varying
the action with respect to the metric we can frame the field
equations in a metric-affine formalism. Using those equations we
can study the cosmological evolution of the theory in detail. Some
specific toy models may help in properly understanding the
dynamics of the theory. We also intend to study the energy
conditions in the background of the theory to put constraints on
the model. The paper is organized as follows: The basic field
equations of $f(Q,T_{\mu\nu}T^{\mu\nu})$ gravity is derived in
section 2. Cosmological evolution of EMSSTG is explored in section
3. Section 4 is dedicated to studying the dynamics of some
specific toy models. In section 5 we investigate the energy
conditions in the background of EMSSTG thus constraining the
theory. Finally the paper ends with some concluding remarks in
section 6.

\section{Field equations of $f(Q,T_{\mu\nu}T^{\mu\nu})$ Gravity}
In this section we will discuss the various geometrical
preliminaries required to frame a relativistic theory of
gravitation and go on to derive the field equations of
$f(Q,T_{\mu\nu}T^{\mu\nu})$ gravity. We start by considering the
action for $f(Q,T_{\mu\nu}T^{\mu\nu})$ gravity as,
\begin{equation}\label{actionfqt2}
\mathcal{S}=\int
\left[\frac{1}{16\pi}f(Q,\textbf{T}^2)+\mathcal{L}_{m}\right]\sqrt{-g}d^{4}x
\end{equation}
Here $g\equiv det(g_{\mu\nu})$ and
$\textbf{T}^2=T_{\mu\nu}T^{\mu\nu}$, where $T_{\mu\nu}$ is the
matter energy-momentum tensor. Moreover $f(Q,\textbf{T}^2)$ is a
function of the non-metricity $Q$ and $\textbf{T}^{2}$ and
$\mathcal{L}_{m}$ is the matter Lagrangian. The non-metricity
tensor $Q_{\lambda\mu\nu}$ is defined as the covariant derivative
of the metric tensor with respect to the Weyl-Cartan connection
$\tilde{\Gamma}^{\lambda}_{~\mu\nu}$ \cite{ecrev},
\begin{equation}\label{nmten1}
-\nabla_{\lambda}g_{\mu\nu}=Q_{\lambda\mu\nu}=-\frac{\partial
g_{\mu\nu}}{\partial
x^{\lambda}}+g_{\nu\sigma}\tilde{\Gamma}^{\sigma}_{~\mu\lambda}+g_{\sigma\mu}\tilde{\Gamma}^{\sigma}_{~\nu\lambda}
\end{equation}
The Weyl-Cartan connection $\tilde{\Gamma}^{\lambda}_{~\mu\nu}$
can be written as a combination of three entities: the Christoffel
symbol $\Gamma^{\lambda}_{~\mu\nu}$, the contortion tensor
$C^{\lambda}_{~\mu\nu}$ and the disformation tensor
$L^{\lambda}_{~\mu\nu}$, in the following way,
\begin{equation}\label{wcconn}
\tilde{\Gamma}^{\lambda}_{~\mu\nu}=\Gamma^{\lambda}_{~\mu\nu}+C^{\lambda}_{~\mu\nu}+L^{\lambda}_{~\mu\nu}
\end{equation}
In the above expression the first term, i.e. the Christoffel
symbol is basically the Levi-Civita connection of the metric
$g_{\mu\nu}$ given by,
\begin{equation}\label{levicivita}
\Gamma^{\lambda}_{~\mu\nu}=\frac{1}{2}g^{\lambda\sigma}\left(\frac{\partial
g_{\sigma\nu}}{\partial x^{\mu}}+\frac{\partial
g_{\sigma\mu}}{\partial x^{\nu}}-\frac{\partial
g_{\mu\nu}}{\partial x^{\sigma}}\right)
\end{equation}
The contortion tensor $C^{\lambda}_{~\mu\nu}$ is obtained from the
torsion tensor $\tilde{\Gamma}^{\lambda}_{~[\mu\nu]}$, which in
turn is defined as,
\begin{equation}\label{torsionten1}
\tilde{\Gamma}^{\lambda}_{~[\mu\nu]}=\frac{1}{2}\left(\tilde{\Gamma}^{\lambda}_{~\mu\nu}-
\tilde{\Gamma}^{\lambda}_{~\nu\mu}\right)
\end{equation}
Using the above definition the contortion tensor is defined as,
\begin{equation}\label{contorsionten1}
C^{\lambda}_{~\mu\nu}=\tilde{\Gamma}^{\lambda}_{~[\mu\nu]}
+g^{\lambda\sigma}g_{\mu\kappa}\tilde{\Gamma}^{\kappa}_{~[\nu\sigma]}+
g^{\lambda\sigma}g_{\nu\kappa}\tilde{\Gamma}^{\kappa}_{~[\mu\sigma]}
\end{equation}
The disformation tensor is given by,
\begin{equation}\label{disformf}
L^{\alpha}_{~\beta\gamma}=-\frac{1}{2}g^{\alpha\lambda}\left(\nabla_{\gamma}~g_{\beta\lambda}+\nabla_{\beta}~g_{\lambda\gamma}-\nabla_{\lambda}~g_{\beta\gamma}\right)
=-\frac{1}{2}g^{\alpha\lambda}\left(Q_{\gamma\beta\lambda}+Q_{\beta\lambda\gamma}-Q_{\lambda\beta\gamma}\right)
\end{equation}

Using this the non-metricity is given as,
\begin{equation}\label{nonmetf}
Q=-g^{\mu\nu}\left(L^{\alpha}_{~\beta\mu}L^{\beta}_{~\nu\alpha}-L^{\alpha}_{~\beta\alpha}L^{\beta}_{~\mu\nu}\right)
\end{equation}
This nonmetricity invariant is by construction equivalent to
negative of the Einstein Lagrangian, when the covariant derivative
reduces to the partial derivative, i.e.,
$\nabla_{\alpha}=\partial_{\alpha}$. This gauge choice is called
the \textit{coincident gauge} and is consistent with symmetric
teleparallel gravity \cite{dsit}. In the connection given in
eqn.(\ref{wcconn}), the contorsion depends on the torsion tensor,
the disformation depends on the non metricity tensor and the
Levi-Civita connection characterizes the curvature. For nonmetric
gravity we have vanishing torsion and curvature, due to which the
connection becomes equal to just the disformation in the
coincident gauge. Though the Levi-Civita connection is eliminated,
curvature continues to play its physical role in the set-up. Now
since we will take a trivially connected geometry we should have
$\tilde{\Gamma}^{\lambda}_{~\mu\nu}=\Gamma^{\lambda}_{~\mu\nu}+L^{\lambda}_{~\mu\nu}=0$.
Furthermore the gravity is realized as a gauge theory of the group
of translations. Now since this is a class of generalizations in
the trivially connected geometry (vanishing connection)  it is
actually teleparallelized in the metric affine gauge theory.

Moreover the Weyl-Cartan torsion tensor $\tau^{\lambda}_{\mu\nu}$
is defined as,
\begin{equation}\label{wctorten}
\tau^{\lambda}_{\mu\nu}=\frac{1}{2}\left(\tilde{\Gamma}^{\lambda}_{~\mu\nu}-
\tilde{\Gamma}^{\lambda}_{~\nu\mu}\right)
\end{equation}
and the Weyl-Cartan curvature tensor may be defined by,
\begin{equation}\label{wccurveten}
\tilde{R}^{\lambda}_{~\mu\nu\sigma}=\tilde{\Gamma}^{\lambda}_{~\mu\sigma,\nu}
-\tilde{\Gamma}^{\lambda}_{~\mu\nu,\sigma}+\tilde{\Gamma}^{\alpha}_{~\mu\sigma}
\tilde{\Gamma}^{\lambda}_{~\alpha\nu}-\tilde{\Gamma}^{\alpha}_{~\mu\nu}
\tilde{\Gamma}^{\lambda}_{~\alpha\sigma}
\end{equation}

We know that the trace of the energy-momentum tensor $T_{\mu\nu}$
is given as,
\begin{equation}\label{traceemt}
T=g^{\mu\nu}T_{\mu\nu}=g_{\mu\nu}T^{\mu\nu}
\end{equation}
The trace of the non-metricity tensor is given by,
\begin{equation}\label{tracenm}
Q_{\beta}=Q_{\beta~~\lambda}^{~~\lambda}~,
~~~~~~~~~~~~~~\tilde{Q}_{\beta}=Q^{\lambda}_{~~\beta\lambda}
\end{equation}
We will bring in the superpotential of the model defined as,

$$P^{\alpha}_{~~\mu\nu}=\frac{1}{4}\left[-Q^{\alpha}_{~~\mu\nu}+2~Q_{(\mu~~\nu)}^{~~\alpha}+Q^{\alpha}g_{\mu\nu}-\tilde{Q}^{\alpha}g_{\mu\nu}-\delta^{\alpha}_{~(\mu}Q_{\nu)}\right]$$
\begin{equation}\label{suppot}
=-\frac{1}{2}L^{\alpha}_{\mu\nu}+\frac{1}{4}\left(Q^{\alpha}-\tilde{Q}^{\alpha}\right)g_{\mu\nu}-\frac{1}{4}\delta^{\alpha}_{~(\mu}Q_{\nu)}
\end{equation}
Now we can use the non-metricity tensor with the superpotential to
obtain the following relation for the non-metricity
\begin{equation}\label{nmetr}
Q=-Q_{\lambda\alpha\beta}P^{\lambda\alpha\beta}=-\frac{1}{4}\left(-Q^{\lambda\beta\rho}Q_{\lambda\beta\rho}+2~Q^{\lambda\beta\rho}Q_{\rho\lambda\beta}-2~Q^{\rho}\tilde{Q}_{\rho}+Q^{\rho}Q_{\rho}\right)
\end{equation}

Now we proceed to derive the field equations. For this, we take
the variation of the action $\mathcal{S}$ in
eqn.(\ref{actionfqt2}) with respect to the metric tensor and
obtain,
\begin{equation}\label{actionvary1}
\delta \mathcal{S}=\int \frac{1}{16\pi}\delta
\left[f(Q,\textbf{T}^{2})\sqrt{-g}\right]d^{4}x+\int \delta
\left[\mathcal{L}_{m}\sqrt{-g}\right]d^{4}x
\end{equation}
Simplifying the above relation we get,
\begin{equation}\label{actionvary2}
\delta \mathcal{S}=\int \frac{1}{16\pi}
\left[-\frac{1}{2}g_{\mu\nu}\delta
g^{\mu\nu}f(Q,\textbf{T}^{2})+f_{Q}\delta
Q+f_{\textbf{T}^{2}}\delta \textbf{T}^{2}-8\pi T_{\mu\nu}\delta
g^{\mu\nu}\right]\sqrt{-g}d^{4}x
\end{equation}
where $f_{Q}=\frac{\partial f}{\partial Q}$ and
$f_{\textbf{T}^{2}}=\frac{\partial f}{\partial \textbf{T}^{2}}$.
Now in the above expression we have made use of the relations,
\begin{equation}\label{varyrootg}
\delta (\sqrt{-g})=-\frac{1}{2}\sqrt{-g}g_{\mu\nu}\delta
g^{\mu\nu}
\end{equation}
and
\begin{equation}\label{varyemt}
T_{\mu\nu}=-\frac{2}{\sqrt{-g}}\frac{\delta(\sqrt{-g}\mathcal{L}_{m})}{\delta
g^{\mu\nu}}
\end{equation}
Considering that $\mathcal{L}_{m}$ depends only on the metric
tensor and not on its derivatives, we get the relation,
\begin{equation}\label{emtten2}
T_{\mu\nu}=g_{\mu\nu}\mathcal{L}_{m}-2\frac{\partial
\mathcal{L}_{m}}{\partial g^{\mu\nu}}
\end{equation}

Now we see from eqn.(\ref{actionvary2}) that we have to calculate
the quantities $\delta Q$ and $\delta \textbf{T}^{2}$. After some
rigorous calculations $\delta Q$ is obtained as,
\begin{equation}\label{varyq}
\delta Q=2P_{\alpha\nu\rho}\nabla^{\alpha}\delta
g^{\nu\rho}-\left(P_{\mu\alpha\beta}Q_{\nu}^{~~\alpha\beta}
-2Q^{\alpha\beta}_{~~\mu}P_{\alpha\beta\nu}\right)\delta
g^{\mu\nu}
\end{equation}

For an explicit calculation of $\delta Q$ see ref.\cite{ftq1}. Now
let us define a quantity $\theta_{\mu\nu}$ as,
\begin{equation}\label{theta}
\theta_{\mu\nu}=\frac{\delta(T_{\alpha\beta}T^{\alpha\beta})}{\delta
g^{\mu\nu}}
\end{equation}
From the above relation we can write $\delta
\textbf{T}^{2}=\delta(T_{\alpha\beta}T^{\alpha\beta})=\theta_{\mu\nu}\delta
g^{\mu\nu}$, where an explicit relation for $\theta_{\mu\nu}$ is
calculated as,
\begin{equation}\label{theta1}
\theta_{\mu\nu}=2T^{\alpha}_{\mu}T_{\nu\alpha}-2\mathcal{L}_{m}\left(T_{\mu\nu}
-\frac{1}{2}g_{\mu\nu}T\right)-TT_{\mu\nu}-4T^{\alpha\beta}\frac{\delta^{2}\mathcal{L}_{m}}{\delta
g^{\mu\nu}\delta g^{\alpha\beta}}
\end{equation}
Using all these in eqn.(\ref{actionvary2}) we get the variation of
the action as,

$$\delta \mathcal{S}=\int \frac{1}{16\pi}
\left[-\frac{1}{2}g_{\mu\nu}\delta
g^{\mu\nu}f(Q,\textbf{T}^{2})+f_{Q}\left\{2P_{\alpha\mu\nu}\nabla^{\alpha}\delta
g^{\mu\nu}-\left(P_{\mu\alpha\beta}Q_{\nu}^{~~\alpha\beta}
-2Q^{\alpha\beta}_{~~\mu}P_{\alpha\beta\nu}\right)\delta
g^{\mu\nu}\right\}\right.$$
\begin{equation}\label{actionvary3}
\left.+f_{\textbf{T}^{2}}\theta_{\mu\nu}\delta g^{\mu\nu}-8\pi
T_{\mu\nu}\delta g^{\mu\nu}\right]\sqrt{-g}d^{4}x
\end{equation}
Now the term $2f_{Q}\sqrt{-g}P_{\alpha\mu\nu}\nabla^{\alpha}\delta
g^{\mu\nu}$ on integration and with the use of boundary condition
gives
$-2\nabla^{\alpha}\left(f_{Q}\sqrt{-g}P_{\alpha\mu\nu}\right)\delta
g^{\mu\nu}$. Using this and equating the variation of the action
to zero we finally obtain the field equations of
$f(Q,\textbf{T}^{2})$ gravity as,
\begin{equation}\label{fieldeq1}
-\frac{2}{\sqrt{-g}}\nabla_{\alpha}\left(f_{Q}\sqrt{-g}P^{\alpha}_{~\mu\nu}\right)
-\frac{1}{2}f(Q,\textbf{T}^{2})g_{\mu\nu}+f_{\textbf{T}^{2}}\theta_{\mu\nu}-f_{Q}\left(P_{\mu\alpha\beta}Q_{\nu}^{~~\alpha\beta}
-2Q^{\alpha\beta}_{~~\mu}P_{\alpha\beta\nu}\right)=8\pi T_{\mu\nu}
\end{equation}
For the special case $f(Q,\textbf{T}^{2})=f(Q)$, these field
equations reduce to those of $f(Q)$ gravity. Moreover for the
choice $f(Q,\textbf{T}^{2})=Q$ the above field equations reduce to
the field equations of the symmetric teleparallel equivalent of
general relativity \cite{stegr1, stegr2, stegr3, stegr4, stegr5}.
The $(1,1)$- form of the field equations are given by,
\begin{equation}\label{11 form}
f_{\textbf{T}^{2}}\theta_{~\nu}^{\mu}-8\pi
T_{~\nu}^{\mu}=\frac{2}{\sqrt{-g}}\nabla_{\alpha}\left(f_{Q}\sqrt{-g}P^{\alpha\mu}_{~~~\nu}\right)
+\frac{1}{2}f(Q,\textbf{T}^{2})\delta_{~\nu}^{\mu}+f_{Q}\left(P^{\mu}_{~~\alpha\beta}Q_{\nu}^{~~\alpha\beta}\right)
\end{equation}
Now we consider two constraints, the torsion tensor
$\tau^{\alpha}_{~\beta\gamma}=0$ and the curvature tensor
$R^{\alpha}_{~\beta\mu\nu}=0$, and using the Lagrange multiplier
method, we find the variation with respect to the connection. For
that we first define the hyper momentum tensor density as
\cite{ftq1},
\begin{equation}\label{hyper}
H_{\rho}^{~\alpha\beta}\equiv
\frac{\sqrt{-g}}{16\pi}f_{\textbf{T}^{2}}\frac{\delta
\textbf{T}^{2}}{\delta
\tilde{\Gamma}^{\rho}_{~\alpha\beta}}+\frac{\delta\sqrt{-g}\mathcal{L}_{m}}{\delta
\tilde{\Gamma}^{\rho}_{~\alpha\beta}}
\end{equation}
where $\tilde{\Gamma}^{\rho}_{~\alpha\beta}$ is the Weyl-Cartan
connection defined in eq.(\ref{wcconn}).

Now using the above expression and taking the variation of the
gravitational action with respect to the connection we get
\cite{ftq1},
\begin{equation}\label{varycon}
\nabla_{\mu}\nabla_{\nu}\left(\sqrt{-g}f_{Q}P^{\mu\nu}_{~~~\alpha}+4\pi
H_{\alpha}^{~~\mu\nu}\right)=0
\end{equation}
Here we could have used the inertial variation by setting the
connection in its pure gauge form in the action. We could have
also used a general connection in the action and supplement it
with Lagrange multipliers to eliminate the curvature and torsion.
We will be using the above quantities in the next section where we
investigate the conservation of the matter energy-momentum tensor.
In case of minimal coupling of matter there is obviously no
violation of the equivalence principle. In that case for the gauge
interpretation of the theory we should have
$\partial_{\mu}\rightarrow \nabla_{\mu}$, whereas the Weitzenbock
teleparallelism will require $\partial_{\mu}\rightarrow D_{\mu}$
\cite{new1}. The gauge choices involved in our theory are quite
compatible with the cosmological evolution. The trivially
connected geometry considered in our theory would mean a universe
devoid of torsion and curvature and the gravitational interaction
will be fully realized from nonmetricity, at least mathematically.
Physically curvature will still have its effects, but due to the
vanishing of the connection, all the inertial effects connected
with the geometry will be absent. The gravitational effects
arising from such a set-up will be interesting. Cosmological
compatibility will not be an issue with such a set-up because
curvature still plays its role physically as discussed earlier.

Before we go on to study other important features, it will be
interesting to discuss the propagating degrees of freedom of the
theory. From a thorough Hamiltonian analysis, we come to know that
Einstein's general relativity has two propagating degrees of
freedom (DOF) in four dimensions. The metric perturbations can be
split into three parts: scalar, vector and tensor components. It
is seen that the scalar and the vector components are not
propagating degrees of freedom but are mere constraints.
Alternatively, we see that in GR we have ten partial differential
equations. Four constraints can be obtained from the Bianchi
identities and we also have four gauge transformations, which are
worth degrees of freedom and are arbitrary. So DOF of GR$=10- 4- 4
= 2$. The metric $f(R)$ gravity which is considered as the
representative of the modified gravity theory has three DOF
\cite{dof1}. For $f(\tau)$ gravity ($\tau$ is torsion scalar) it
is found that due to the violation of local Lorentz invariance
there are five DOF \cite{dof2}. It is also shown by the authors in
\cite{dof2} that in general for $f(\tau)$ gravity there are
$(D-1)$ extra degrees of freedom in $D$ dimensions when compared
to Einstein gravity. For $f(Q)$ gravity it was shown in
\cite{dof3} that it respects local Lorentz symmetry and harbours
no extra degrees of freedom compared to GR. So due to the absence
of any extra polarization modes $f(Q)$ gravity also has two
degrees of freedom. Probably this is due to the fact that STG and
$f(Q)$ gravity does not rely on curvature or torsion for its
formulation. Instead, the whole theory is developed using
non-metricity, which does not produce extra degrees of freedom.
Now any modification in the source term (matter sector
$T_{\mu\nu}$) in principle is not expected to bring about any
change in the propagating degrees of freedom even if it is
included in the gravitational Lagrangian. This is because it will
involve terms comprising of pressure $p$, matter density $\rho$
and its derivatives, that do not participate in determining the
propagating degrees of freedom. Hence theories like $f(Q,T)$
should possess similar degrees of freedom as $f(Q)$ theory. Using
similar arguments we can say that $f(Q,\mathbf{T}^{2})$ gravity
theory will also have two propagating degrees of freedom in four
dimensions. In fact we need to perform a detailed metric
perturbation analysis to check that unlike GR if any vector
perturbation modes become dynamical degrees of freedom for
$f(Q,\mathbf{T}^2)$ gravity theory to be sure about our arguments.
This can be attempted in a future project.

\subsection{Conservation of matter energy-momentum tensor \& the momentum conservation equation}
The covariant derivative of a $(1,1)$-form tensor $w^{\mu}_{~\nu}$
may be expressed as,
\begin{equation}\label{covder}
\nabla_{\mu}
w^{\mu}_{~\nu}=D_{\mu}w^{\mu}_{~\nu}-\frac{1}{2}Q_{\lambda}w^{\lambda}_{~\nu}
-L^{\lambda}_{~\mu\nu}w^{\mu}_{~\lambda}
\end{equation}
where $D_{\mu}$ represents the covariant derivative with respect
to the Levi-Civita connection $(\Gamma^{\alpha}_{~\mu\nu})$
defined as
$\Gamma^{\alpha}_{~\mu\nu}=\tilde{\Gamma}^{\alpha}_{~\mu\nu}-L^{\alpha}_{~\mu\nu}$.

Using the above relation we take the covariant derivative of the
field equations (\ref{11 form}) in (1,1)-form and get,
\begin{equation}\label{emtcons1}
D_{\mu}\left[f_{\textbf{T}^2}\theta^{\mu}_{~\nu}-8\pi
T^{\mu}_{~\nu}\right]+\frac{8\pi}{\sqrt{-g}}\nabla_{\alpha}\nabla_{\mu}H_{\nu}^{~\alpha\mu}
=\frac{1}{2}f_{\textbf{T}^2}\partial_{\nu}\textbf{T}^{2}
+\frac{1}{\sqrt{-g}}Q_{\mu}\nabla_{\alpha}\left(f_{Q}\sqrt{-g}P^{\alpha\mu}_{~~~\nu}\right)
\end{equation}
where $\partial_{\nu}$ represents partial derivative with respect
to the coordinate $x^{\nu}$. From the above relation we see that,
\begin{equation}\label{emtcons2}
D_{\mu}T^{\mu}_{~\nu}=\frac{1}{8\pi}\left[D_{\mu}\left(f_{\textbf{T}^2}
\theta^{\mu}_{~\nu}\right)+\frac{8\pi}{\sqrt{-g}}\nabla_{\alpha}
\nabla_{\mu}H_{\nu}^{~\alpha\mu}-\frac{1}{\sqrt{-g}}Q_{\mu}\nabla_{\alpha}\left(f_{Q}\sqrt{-g}
P^{\alpha\mu}_{~~~\nu}\right)-\frac{1}{2}f_{\textbf{T}^2}\partial_{\nu}\textbf{T}^{2}\right]\neq0
\end{equation}
So we can see that the matter energy-momentum tensor is not
conserved in this case. Now we may introduce a tensor
$A^{\nu}_{~\alpha}$ to solve eqn.(\ref{varycon}) such that
\cite{ftq1},
\begin{equation}\label{newten}
\nabla_{\mu}\left(\sqrt{-g}f_{Q}P^{\mu\nu}_{~~~\alpha}+4\pi
H_{\alpha}^{~~\mu\nu}\right)=\sqrt{-g}A^{\nu}_{~\alpha}
\end{equation}
After some rigorous calculations we reach the following expression
for the covariant derivative of the energy-momentum tensor,

$$D_{\mu}T^{\mu}_{~\nu}=\frac{1}{8\pi}\left[D_{\mu}\left(f_{\textbf{T}^2}
\theta^{\mu}_{~\nu}\right)+\frac{16\pi}{\sqrt{-g}}\nabla_{\alpha}
\nabla_{\mu}H_{\nu}^{~\alpha\mu}-8\pi
\nabla_{\mu}\left(\frac{1}{\sqrt{-g}}\nabla_{\alpha}H_{\nu}^{~\alpha\mu}\right)
+2\nabla_{\mu}A^{\mu}_{~\nu}\right.$$
\begin{equation}\label{emtcons3}
\left.-\frac{1}{2}f_{\textbf{T}^2}\partial_{\nu}\textbf{T}^{2}\right]
=B_{\nu}\neq0
\end{equation}
As mentioned above the energy-momentum tensor is not conserved for
this theory and we get $D_{\mu}T^{\mu}_{~\nu}=B_{\nu}\neq0$. The
non-conservation vector $B_{\nu}$ is a function of the dynamical
variables $Q$, $\textbf{T}^{2}$ and other thermodynamic parameters
like energy density and pressure. From eqn.(\ref{covder}), it can
be seen that the covariant derivative $\nabla_{\mu}$ is a function
of the covariant derivative on Levi-Civita connection $D_{\mu}$,
trace of the non-metricity tensor $Q_{\lambda}$ and the
disformation tensor $L^{\lambda}_{~\mu\nu}$. So straightforward
algebra will tell us that the $D_{\mu}$ can be expressed as a
function or combination of $\nabla_{\mu}$, $Q_{\lambda}$ and
$L^{\lambda}_{~\mu\nu}$. Now since the theory contains deformation
terms with respect to $D_{\mu}$, it is expected to give
deformation terms with respect to $\nabla_{\mu}$ also. This is
because when one takes derivative with respect to $\nabla_{\mu}$,
invariably one has to go through the derivatives with respect to
$D_{\mu}$ because of the dependence given in eqn.(\ref{covder}).
Unless there is loss of terms due to the effect of the other two
factors in the equation, namely $Q_{\lambda}$ and
$L^{\lambda}_{~\mu\nu}$ (which seems unlikely) the quantity
$\nabla_{\mu}T^{\mu}_{\nu}$ will not be conserved because the
quantity $D_{\mu}T^{\mu}_{\nu}$ is not conserved.

On a general note dissipative processes are not compatible with
Cosmic microwave background radiation (CMBR) or Large scale
structure (LSS). The authors in \cite{new2} studied the
cosmological and solar system consequences of a class of matter
coupling models with geometry. They found that the models
considered generally have some inconsistent behaviour when
compared with the observational data. There is a fair possibility
of this behaviour being transmitted and magnified when we consider
cosmology at the galactic and extra-galactic level. It is expected
that there will be a fair amount of incompatibility with CMBR or
LSS. But it is seen that, this is a purely model dependent
phenomenon. By fine tuning the model parameters it is possible to
get rid of some or all the inconsistencies as clearly worked out
in \cite{new2}. At the large scales (galactic and extra galactic
levels) there are some implications of the non-minimal matter
coupling with geometry. The flattening of the galaxy rotation
curves as a dynamically generated effect can be attributed to this
non-minimal coupling between matter and geometry \cite{new3}. Due
to the non-conservation of the energy-momentum tensor, a deviation
from the geodesic motion sets in. This accounts for the observed
discrepancy between the measured rotation velocity and the
classical prediction. It can also be shown that a special type of
non-minimal matter coupling with geometry can mimic the dark
matter component of the galaxy clusters. For this purpose the
authors of \cite{new4} investigated the Abell cluster A586, which
is a massive nearby relaxed cluster of galaxies in virial
equilibrium. Then one can easily extend the dark matter mimicking
phenomenon to a large sample of galaxy clusters. Dissipative
processes also have their special role in the evolution of radio
galaxies as discussed in \cite{new5}.

If the matter content is described by a perfect fluid with density
$\rho$ and pressure $p$ and the energy-momentum tensor given by
$T^{\mu}_{~\nu}=\left(\rho+p\right)u_{\nu}u^{\mu}+p\delta^{\mu}_{\nu}$,
then following \cite{dsit} we obtain the energy balance equation
as,
\begin{equation}\label{embeq}
\dot{\rho}+3H\left(\rho+p\right)=B_{\alpha}u^{\alpha}
\end{equation}
where dot $(.)$ denotes derivative with respect to time. It can be
clearly seen that this is not the standard continuity equation,
but contains additional deformation terms on the RHS. Here
$B_{\alpha}u^{\alpha}$ is the source term that correlates with the
creation or annihilation of energy. If $B_{\alpha}u^{\alpha}=0$,
then we have an energy conserved gravitational system, otherwise
transfer of energy prevails and particle production takes place in
the system. The momentum conservation equation representing the
motion of massive particles is given by \cite{dsit, ftq1},
\begin{equation}\label{momcons}
\frac{d^{2}x^{\mu}}{ds^{2}}+\Gamma^{\mu}_{\lambda\sigma}u^{\lambda}u^{\sigma}
=\frac{\Upsilon^{\mu\nu}}{\rho+p}\left(B_{\nu}-D_{\nu}p\right)=F^{\mu}
\end{equation}
where $\Upsilon^{\mu\nu}$ is the projection operator given by
$\Upsilon^{\mu\nu}=g^{\mu\nu}+u^{\mu}u^{\nu}$. The term
$F^{\mu}=\frac{\Upsilon^{\mu\nu}}{\rho+p}\left(B_{\nu}-D_{\nu}p\right)$
on the RHS is the deformation term representing additional force
coming from the coupling effects of $Q$ and $\textbf{T}^{2}$. As a
result the dynamical evolution of the massive particles is
non-geodesic in nature. Due to the presence of the projection
operator, this extra force acts orthogonal to the 4-velocity
vector such that $F^{\mu}u_{\mu}=0$. It is known that the
components of the 4-force which are orthogonal to the 4-velocity
of a particle can only affect its motion and the path followed. In
this sense the extra force $F^{\mu}$ derived from the theory is
perfectly physical. Now we write the extra force $F^{\mu}$ as a
combination of three terms as below,
\begin{equation}\label{3break}
F^{\mu}=-\frac{\Upsilon^{\mu\lambda}D_{\lambda}p}{\rho+p}+F^{\mu}_{H}+F^{\mu}_{\mathbf{T}^2}
\end{equation}
where $F^{\mu}_{H}$ and $F^{\mu}_{\mathbf{T}^2}$ are given as,
\begin{equation}\label{breakhyp}
F^{\mu}_{H}=\frac{\Upsilon^{\mu\nu}}{\rho+p}\left[\frac{2}{\sqrt{-g}}\nabla_{\alpha}
\nabla_{\beta}H_{\nu}^{~\alpha\beta}-
\nabla_{\beta}\left(\frac{1}{\sqrt{-g}}\nabla_{\alpha}H_{\nu}^{~\alpha\beta}\right)
+\frac{\nabla_{\alpha}A^{\alpha}_{~\nu}}{4\pi}\right]
\end{equation}

\begin{equation}\label{breakt}
F^{\mu}_{\mathbf{T}^2}=\frac{\Upsilon^{\mu\nu}}{8\pi\left(\rho+p\right)}\left[D_{\alpha}\left(f_{\textbf{T}^2}
\theta^{\alpha}_{~\nu}\right)-\frac{1}{2}f_{\textbf{T}^2}\partial_{\nu}\textbf{T}^{2}\right]
\end{equation}

In eqn.(\ref{3break}) the first term on the right hand side arises
from the typical general relativistic contribution of the pressure
gradient. The second term on the right hand side $F^{\mu}_{H}$ is
the hyper-force whose value is given in eqn.(\ref{breakhyp}). The
third term $F^{\mu}_{\mathbf{T}^2}$ is the extra force mainly
coming from matter, which is given in eqn.(\ref{breakt}). If
$f_{\textbf{T}^2}=0$, it is evident that the extra force coming
from the component $F^{\mu}_{\mathbf{T}^2}$ vanishes. We discuss
the effect of considering a perfect fluid on this extra force in
the next subsection.

\subsection{Matter as perfect fluid}

Now we need to consider a specific type of matter component to
further simplify the field equations. We assume that the universe
is filled with a perfect fluid described by the energy-momentum
tensor,
\begin{equation}\label{emtperfect}
T_{\mu\nu}=\left(\rho+p\right)u_{\mu}u_{\nu}+pg_{\mu\nu}
\end{equation}
where $\rho$ is the energy density and $p$ is the pressure.
Moreover $u_{\mu}$ is the four-velocity of the fluid, which is
normalized as $u_{\mu}u^{\mu}=-1$. Using the above expression we
get,
\begin{equation}\label{Tsquare}
T_{\mu\nu}T^{\mu\nu}=\textbf{T}^2=\rho^{2}+3p^2
\end{equation}
We consider the matter Lagrangian $\mathcal{L}_{m}=p$. From this
assumption we see that the final term of $\theta_{\mu\nu}$ in
eqn.(\ref{theta1}) vanishes because pressure $p$ does not depend
on the metric tensor. Now using eqn.(\ref{emtperfect}) in
eqn.(\ref{theta1}) we get,
\begin{equation}\label{theta2}
\theta_{\mu\nu}=-\left(\rho^{2}+4p\rho+3p^{2}\right)u_{\mu}u_{\nu}
\end{equation}
Using the above expressions the field equations take the following
form,
$$-\frac{2}{\sqrt{-g}}\nabla_{\alpha}\left(f_{Q}\sqrt{-g}P^{\alpha}_{~\mu\nu}\right)
-\frac{1}{2}f(Q,\textbf{T}^{2})
g_{\mu\nu}-f_{Q}\left(P_{\mu\alpha\beta}Q_{\nu}^{~~\alpha\beta}
-2Q^{\alpha\beta}_{~~\mu}P_{\alpha\beta\nu}\right)$$
\begin{equation}\label{fieldeqfinal}
=\left[8\pi\left(\rho+p\right)
+f_{\textbf{T}^{2}}\left(\rho^{2}+4p\rho+3p^{2}\right)\right]u_{\mu}u_{\nu}
+8\pi p g_{\mu\nu}
\end{equation}

Using eqns.(\ref{Tsquare}) and (\ref{theta2}) in
eqn.(\ref{breakt}) we get the component of extra force for perfect
fluid as,
\begin{equation}\label{perfforce}
F^{\mu}_{\mathbf{T}^2}=\frac{\Upsilon^{\mu\nu}}{8\pi\left(\rho+p\right)}\left[D_{\alpha}\left(-f_{\textbf{T}^2}
\left(\rho^{2}+4p\rho+3p^{2}\right)u^{\alpha}u_{\nu}\right)-\frac{1}{2}f_{\textbf{T}^2}\partial_{\nu}\left(\rho^{2}+3p^2\right)\right]
\end{equation}
If we consider the equation of state (EoS) of matter as $p=k\rho$,
where $k$ is the EoS parameter, then for $k=-1$ ($\Lambda$CDM) and
$k=-1/3$, the first term inside the bracket of the above equation
vanishes. For low energy density regime ($\rho^2\rightarrow 0$),
both the terms inside the bracket vanish and hence the total force
component $F^{\mu}_{\mathbf{T}^2}$ vanishes. This should
correspond to very late universe which is highly accelerating in
nature and dominated by dark components. But in the early universe
($\rho^2\rightarrow \infty$), and this force component
($F^{\mu}_{\mathbf{T}^2}$) dominates the first two components in
eqn.(\ref{3break}). So this force component has its roots in the
quantum fluctuations of the early universe. This also shows that
the total additional force $F^{\mu}$ has significant effect in the
evolution of the early universe, but slowly fades away to give
general relativistic effects in the late universe.

We know that modified gravity theories are considered equivalent
to the exotic matter components or dark energy that is considered
responsible for driving the late cosmic acceleration. In our model
we are talking of a highly non standard form of matter coupling
with geometry in the action. Since the matter content creates the
geometry of the spacetime, it is expected that the matter in such
a theory will not be of the usual nature, but will have exotic
properties as mentioned above. The major exotic property will be
the negative pressure that is possessed by the dark energy models
which enables them to drive the cosmic acceleration. It may have
other exotic features, which are subject to testing, detection and
further research. So here we are dealing with a cosmic fluid with
mysterious properties of its own and probably comparison with
usual matter will give highly contrasting results. We know that
even though dark energy and dark matter have exotic properties,
they do satisfy the standard continuity equation.  But for our
model we can see that due to the non-minimal matter coupling the
standard continuity equation is not satisfied. This tells us that
both the matter component and the force acting on it in the
gravity field are highly non standard. So it can be stated that,
standard matter creates a standard spacetime geometry in GR, a non
standard matter in our theory creates a non standard geometry of
spacetime resulting in non geodesic motion and extra force.

As we have already seen that in the nonmetric formulation the
connection is totally trivialized
$(\tilde{\Gamma}^{\alpha}_{~\mu\nu}=0)$ by means of a
diffeomorphism and as a result the inertial connection vanishes
from the nonmetricity sector. This formulation is thus a subtle
improvement of GR, since the minimally coupled fermions are still
connected metrically \cite{new6}. Moreover since the pure gravity
sector is now trivially connected, effectively nothing changes but
just the higher-derivative boundary term disappears from the
action. Non-minimal matter coupling to curvature based theories
have been widely studied in literature \cite{new7, new8, new9}.
Energy momentum squared gravity (EMSG) is a class of theories
where matter coupling is introduced in the form of $T^2$ with the
scalar curvature $R$ \cite{emsgorg, lqg1, new10}. Such theories
have a lot of salient and interesting features. However matter
coupling to curvature does have its own problems. Due to the
higher derivative property of the curvature scalar $R$, these
theories are best described as effective theories, which can give
rise to problems at certain limits \cite{dsit}. A good example of
this will be, for the density of a canonical scalar field $\phi$
the non-minimal coupling of the form $f(R)L_\phi$ ($L_\phi$ is the
Lagrangian corresponding to the scalar field) introduces a kinetic
term which does not fit into the viable Horndeski class. However
it should be mentioned that such problems are expected to
disappear when the coupling is formulated in the metric-affine
approach because the field equations remain in the second order
\cite{new11}. Thus it is worthy to consider nonminimal matter
couplings with non-metricity Q, because the scalar invariant
involves no higher derivatives. So a coupling of the form
$f(Q)T^2$ or $f(Q)+T^2$ or any other forms involving $Q$ and $T^2$
results in second order equations of motion. So the models that
will be considered here are inspired by the well studied forms of
curvature matter couplings in the literature, but here the
curvature will be simply replaced by the non-metricity. The
motivation is to see whether the subtle improvement of the
geometrical formulation of $f(Q)$ gravity, when implemented in the
matter sector, would allow more universally consistent and viable
realizations of the nonminimal curvature-matter coupling theories
\cite{dsit}. This is the basic advantage of considering
non-metricity in place of curvature scalar.

\section{Cosmology of EMSSTG}
In this section we intend to explore the cosmological applications
of the above derived theory. We will assume that the universe is
described by the homogeneous, isotropic and spatially flat
Friedmann-Lemaitre-Robertson-Walker (FLRW) metric given by,
\begin{equation}\label{flrwmet}
ds^{2}=-dt^{2}+a(t)^{2}\left(dx^{2}+dy^{2}+dz^{2}\right)
\end{equation}
where $a(t)$ is the cosmological scale factor representing the
expansion of the universe. The expansion rate may be given by the
Hubble parameter $H$ defined as,
\begin{equation}\label{hubble}
H\equiv \frac{\dot{a}}{a}
\end{equation}
where dot (.) represents derivative with respect to time. From
some rigorous calculations we see that for the above FLRW metric
the non-metricity $Q$ may be given in terms of Hubble parameter as
\cite{dsit, ftq1},
\begin{equation}\label{nmflrw}
Q=6H^{2}
\end{equation}

Using the metric (\ref{flrwmet}) we get the modified Friedmann
equations for the $f(Q,\textbf{T}^{2})$ gravity as,
\begin{equation}\label{mflrw1}
6f_{Q}H^{2}-\frac{1}{2}f(Q,\textbf{T}^{2})=8\pi
\rho+f_{\textbf{T}^{2}}\left(\rho^{2}+4p\rho+3p^{2}\right)
\end{equation}
\begin{equation}\label{mflrw2}
6f_{Q}H^{2}-\frac{1}{2}f(Q,\textbf{T}^{2})-2\left(\dot{f_{Q}}H+f_{Q}\dot{H}\right)=-8\pi
p
\end{equation}
where as mentioned earlier dot (.) represents derivative with
respect to time. It should be noted that we have derived the above
FLRW equations for a universe filled with perfect fluid with the
energy-momentum tensor given by eqn.(\ref{emtperfect}). In the
first FLRW equation we see that there are two correction terms
$1/2 f$ and
$f_{\textbf{T}^{2}}\left(\rho^{2}+4p\rho+3p^{2}\right)$ when
compared to standard equations of GR. The first term is present in
the other extensions of symmetric teleparallel gravity like the
$f(Q)$ and $f(Q,T)$ gravity. But the second term is the one to
look out for, because this is the term that is characteristic of
$f(Q,\textbf{T}^{2})$ theory. Moreover we see that the equations
follow perfect correspondence and reduce to those of STG, $f(Q)$
or $f(Q,T)$ under suitable limits.  We can write the FLRW
equations in the standard form for general relativity as below,
\begin{equation}\label{flrwstan1}
3H^{2}=8\pi \rho_{eff}
\end{equation}
\begin{equation}\label{flrwstan2}
2\dot{H}+3H^{2}=-8\pi p_{eff}
\end{equation}
where
\begin{equation}\label{rhoeff}
\rho_{eff}=\frac{1}{16\pi
f_{Q}}\left[8\pi\rho+\frac{1}{2}f(Q,\textbf{T}^{2})+f_{\textbf{T}^{2}}\left(\rho^{2}
+4p\rho+3p^{2}\right)\right]
\end{equation}
and
\begin{equation}\label{peff}
p_{eff}=\frac{1}{8\pi f_{Q}}\left[-8\pi
p+\frac{1}{2}f(Q,\textbf{T}^{2})-9f_{Q}H^{2}+2\dot{f_{Q}}H \right]
\end{equation}
Now the effective energy density and pressure will follow the
standard continuity equation given as,
\begin{equation}\label{contstan}
\dot{\rho}_{eff}+3H\left(\rho_{eff}+p_{eff}\right)=0
\end{equation}
The effective equation of state $\omega_{eff}$ of the dark energy
derived from the theory may be given by,
\begin{equation}\label{effeos}
\omega_{eff}=\frac{p_{eff}}{\rho_{eff}}=\frac{2\left[-8\pi
p+\frac{1}{2}f(Q,\textbf{T}^{2})-9f_{Q}H^{2}+2\dot{f_{Q}}H
\right]}{\left[8\pi\rho+\frac{1}{2}f(Q,\textbf{T}^{2})+f_{\textbf{T}^{2}}\left(\rho^{2}
+4p\rho+3p^{2}\right)\right]}
\end{equation}
To study the late cosmic acceleration of universe we have devised
a parameter known as the deceleration parameter $q$, which is
defined as,
\begin{equation}\label{decp}
q=-1-\frac{\dot{H}}{H^{2}}=\frac{1}{2}\left(1+3\omega_{eff}\right)
\end{equation}
Using eqn.(\ref{effeos}) in eqn.(\ref{decp}) we can write the
deceleration parameter for this theory as,
\begin{equation}\label{decpfqt2}
q=\frac{1}{2}\left[1+\frac{6\left[-8\pi
p+\frac{1}{2}f(Q,\textbf{T}^{2})-9f_{Q}H^{2}+2\dot{f_{Q}}H
\right]}{\left[8\pi\rho+\frac{1}{2}f(Q,\textbf{T}^{2})+f_{\textbf{T}^{2}}\left(\rho^{2}
+4p\rho+3p^{2}\right)\right]}\right]
\end{equation}
A negative value of this parameter corresponds to the accelerated
expansion of the universe. We will study this parameter in detail
for the specific models.

\subsection{The de-Sitter solution}
Now let us explore the vacuum solutions, and correspondingly check
that whether our theory admits any de-Sitter like solution or not.
To perform that we need to constrain our field equations by
considering $\rho=p=0$ and $H=H_{0}$ (a constant). Using these in
eqns.(\ref{mflrw1}) and (\ref{mflrw2}) we respectively obtain,
\begin{equation}\label{desit1}
6f_{Q}H_{0}^{2}-\frac{1}{2}f=0
\end{equation}
and
\begin{equation}\label{desit2}
6f_{Q}H_{0}^{2}-\frac{1}{2}f-2H_{0}\dot{f_{Q}}=0
\end{equation}
Using the above two equations we get $f_{Q}=f_{0}$ (a constant),
which on integration gives $f=f_{0}Q+\epsilon$ where $\epsilon$ is
the constant of integration. Finally using this result for $f$ in
the eqn.(\ref{desit1}) we get,
\begin{equation}\label{hubdesit}
H_{0}=\sqrt{\frac{\epsilon}{6f_{0}}}
\end{equation}
which is similar to the results obtained in \cite{ftq1, dsit}. For
$f_{0}=1$, we get $H_{0}=\sqrt{\frac{\epsilon}{6}}$, which is
equivalent to the result obtained for general relativity. We know
that for GR, $H_{0}\propto \sqrt{\Lambda}$, where $\Lambda$ is the
cosmological constant. From this we can see that the constant of
integration $\epsilon$ plays the role analogous to the
cosmological constant. So it is clear that our EMSSTG theory
admits de-Sitter like evolution of the universe when subjected to
constraints corresponding to vacuum. Using the relation
(\ref{decp}) it is quite straightforward to get ideas about the
deceleration parameter $q$ and the effective EoS parameter
$\omega_{eff}$ for the de-Sitter universe. Using the derived value
of the Hubble parameter from (\ref{hubdesit}) in eqn.(\ref{decp}),
we get $q=-1$, which is the expected value for de-Sitter universe.
Similarly the effective EoS parameter is obtained as
$\omega_{eff}=-1$, which mimics the cosmological constant. So the
universe in the absence of matter is dominated by vacuum energy
that drives the accelerated expansion of the universe. This is in
accordance with the known picture of the evolution of the universe
and the corresponding literature.

\section{Some specific Toy-models}
We basically consider two specific toy-models in this section. The
models are chosen depending on the nature of coupling between $Q$
and $\textbf{T}^2$. In the first models we will consider an
additive form and in the second model we will consider a product
form. Both the models should have well-defined STG limits, which
is important for correspondence. We will derive the FLRW equations
and explore the cosmological evolution of the models.

\subsection{Model: 1}
Now we need to specify a particular form of $f(Q,\textbf{T}^2)$ to
further explore the field equations. We will assume the general
form
\begin{equation}\label{toymodel}
f(Q,T_{\mu\nu}T^{\mu\nu})=f(Q,\textbf{T}^2)=Q+\eta
\left(T_{\mu\nu}T^{\mu\nu}\right)^{n}=Q+\eta
\left(\textbf{T}^2\right)^{n}
\end{equation}
For $\eta\rightarrow 0$ we recover the STG limit of the theory.
Using eqns.(\ref{emtperfect}), (\ref{Tsquare}), (\ref{theta2}) and
(\ref{toymodel}) in eqn.(\ref{fieldeq1}) we get simplified field
equations as,

$$-\frac{2}{\sqrt{-g}}\nabla_{\alpha}\left(\sqrt{-g}P^{\alpha}_{~~\mu\nu}\right)
-\frac{1}{2}Q
g_{\mu\nu}-\left(P_{\mu\alpha\beta}Q_{\nu}^{~~\alpha\beta}
-2Q^{\alpha\beta}_{~~~\mu}P_{\alpha\beta\nu}\right)$$
\begin{equation}\label{simplefieldeq}
=\left[8\pi\left(\rho+p\right)
+n\eta\left(\rho^{2}+3p^{2}\right)^{n-1}\left(\rho^{2}+4p\rho+3p^{2}\right)\right]u_{\mu}u_{\nu}
+\left[8\pi
p+\frac{\eta}{2}\left(\rho^{2}+3p^{2}\right)^{n}\right]g_{\mu\nu}
\end{equation}
The corresponding Friedmann equations become,
\begin{equation}\label{flrwmod1}
3H^{2}=8\pi
\rho+\eta\left(\rho^{2}+3p^{2}\right)^{n-1}\left[\left(n+\frac{1}{2}\right)\left(\rho^{2}+3p^{2}\right)+4np\rho\right]
\end{equation}
\begin{equation}\label{flrwmod2}
3H^{2}-2\dot{H}=\frac{\eta}{2}\left(\rho^{2}+3p^{2}\right)^{n}-8\pi
p
\end{equation}
From the above equations we see that the correction terms are
higher order terms in the density $\rho$ which should dominate in
the high energy density regime as $\rho\rightarrow\infty$, i.e.
early universe. In the later universe these terms gradually fade
away giving the effects of the standard FLRW model. Looking at the
first Friedmann equation we see that the correction terms are
somewhat similar in form to those arising from quantum gravity
effects in Loop quantum gravity \cite{lqg}. These equations are
also comparable to those arising from the braneworld models
\cite{brane}. So we are interested in searching for solutions of
bouncing cosmology. We consider the following barotropic equation
of state for the fluid
\begin{equation}\label{barotropic}
p=k\rho
\end{equation}
where $k$ is the barotropic parameter. Using this in
eqn.(\ref{flrwmod1}) we see that the correction terms are
quadratic in energy density $\rho$. At small energy density regime
$\rho^{2}\approx 0$ and we recover the standard Friedmann equation
for $n\geq 1$. At high energy densities (early universe) we see
that for $n=1$, we have two critical points i.e. $H=0$ at,
\begin{equation}\label{criticalrho}
\rho_{c_1}=0~,~~~~~~~~~~~~~~~\rho_{c_2}=-\frac{16\pi}{\eta\left(9k^{2}+8k+3\right)}
\end{equation}
The first critical point is quite expected and corresponds to an
empty universe, where the expansion is supposed to be driven by
vacuum energy. The second critical density is more interesting and
makes sense only if ~~$\eta\left(9k^{2}+8k+3\right)<0$
$\Rightarrow$ $\eta<0$. So for negative values of $\eta$ we have a
cosmological bounce at a finite non-zero value of energy density.
For an early radiation dominated universe i.e. for $k=1/3$, we see
that the bounce occurs at $\rho_{c}=-\frac{12\pi}{5\eta}$. At this
point one can easily check that $\dot{H}=-\frac{176
\pi^{2}}{25\eta}>0$ for $\eta<0$. This indicates that the early
universe does not start from an initial singularity but there is a
possibility of a cut-off energy density ($\rho_{c}$) at which the
universe undergoes bounce. So for this model the cosmological
singularity problem can be solved by considering suitable initial
conditions. We fortunately get an explicit solution of the
eqns.(\ref{flrwmod1}) and (\ref{flrwmod2}) for $n=1$ and $k=-1$ as
below,
\begin{equation}\label{soln1}
a(t)=e^{Ct}~,~~~~~~~\rho(t)=\frac{\sqrt{2\left(3C^{2}\eta+8\pi^2\right)}-4\pi}{2\eta}
\end{equation}
where $C$ is the integration constant. This solution resembles
with the solution of the de-Sitter scenario but in the presence of
a non-vanishing matter energy-momentum tensor. Note that here we
have made use of the initial condition $a(0)=1$. Here the energy
density obtained is constant which can mimic an empty universe in
some suitable limits of the parameters (say
$\eta\rightarrow\infty$). For such a scenario the deceleration
parameter will have a constant negative value throughout the
evolution of the universe. So this solution does not generate much
cosmological interest.

Now from the eqn.(\ref{contstan}) using eqns.(\ref{rhoeff}),
(\ref{peff}) and (\ref{toymodel}) we get the continuity equation
for this model as,
\begin{equation}\label{toycont1}
\dot{\rho}=-\frac{3H\left[-18H^{2}+4\dot{H}+16\left(1-2k\right)\pi\rho
+\eta\left(1+3k^{2}\right)^{n-1}\left(3+2n+k\left(9k+8n+6kn\right)\right)\rho^{2n}\right]}
{2\left[8\pi+\eta\left(1+3k^{2}\right)^{n-1}n\left(1+2n+8kn+k^{2}\left(3+6n\right)\right)
\rho^{2n-1}\right]}
\end{equation}
Further, for $n=1$, the above equation reduces to,
\begin{equation}\label{toycont2}
\dot{\rho}+3H\left(\rho+p\right)=\frac{3H\left[18H^{2}-4\dot{H}+\rho\left(48k\pi+\eta
\left(1+k\left(14+k\left(19+18k\right)\right)\right)\rho\right)\right]}
{16\pi+2\eta\left(3+k\left(8+9k\right)\right)\rho}
\end{equation}
The non-conservation term can clearly be seen on the RHS of the
above equation. For matter energy-momentum tensor conservation we
should have
\begin{equation}\label{toycont3}
\rho\left[48k\pi+\eta
\left(1+k\left(14+k\left(19+18k\right)\right)\right)\rho\right]=4\dot{H}-18H^{2}
\end{equation}
It is obvious looking at the above conservation equations
(\ref{toycont1}) and (\ref{toycont2}), it is not straightforward
to integrate them and get a general solution. But for $k=-1$ and
$n=1$ we can integrate the continuity equations and get
expressions similar to eqn.(\ref{soln1}). Using $k=-1$ and the
energy density given in eqn.(\ref{soln1}) in the above condition
for energy momentum conservation (\ref{toycont3}) we get the
following solution for the scale factor,
\begin{equation}\label{toycont4}
a(t)=\frac{B}{\cosh\left[C\left(\frac{9t}{2}-A\right)\right]^{2/9}}
\end{equation}
where $A$ and $B$ are new integration constants and the above
expression suggests that if the evolution of the universe follows
the above scale factor then the matter energy-momentum tensor is
conserved for the model. Obviously it should be kept in mind that
this solution is valid only around the $\Lambda CDM$ regime. Using
the expression for scale factor in eqn.(\ref{toycont4}) we get the
Hubble parameter as,
\begin{equation}\label{hubmod1}
H(t)=-C\tanh\left[C\left(\frac{9t}{2}-A\right)\right]
\end{equation}
Using the above expression for Hubble parameter in
eqn.(\ref{decp}) we get the expression for the deceleration
parameter as,
\begin{equation}
q(t)=-1+\frac{9}{2}~Cosech\left[C\left(\frac{9t}{2}-A\right)\right]^2
\end{equation}

In order to obtain a dimensionless form of the evolution
parameters we introduce the following transformations
\cite{dimless},
\begin{equation}\label{dimless1}
H=H_{0}h, ~~~~~~t=\frac{\tau_p}{H_{0}}, ~~~~~~\rho=3H_{0}^{2}r
\end{equation}
where $\left(h, \tau_p, r\right)$ is a set of dimensionless
variables, and $H_{0}$ is a fixed value of the Hubble function,
which is generally taken as its present value setting the scale of
the universe at present time.

In order to get greater insight into the above obtained form of
scale factor we generate a plot of $a$ against the dimensionless
time $\tau_p$ in figure (1). From the figure we see that the scale
factor grows with time, which complies with our observation of an
expanding universe. So the conservation of matter energy momentum
may be a possibility for this model subject to proper initial
conditions. Moreover we see that in the late universe the
dependency of the scale factor on the initial conditions become
just a little more prominent compared to the early universe. In
fig.(2) we have generated plots for the above Hubble function $h$
for different initial conditions against dimensionless time
$\tau_p$ to gain greater insight into the model. From the plot we
see that the Hubble function remains at the positive level, which
is required for an expanding universe. Gradually it decays over
time, especially in the late universe, which is expected. In
figure (3) we have plotted the deceleration parameter $q$ against
dimensionless time $\tau_p$ for different initial conditions. We
see that there is a smooth transition from positive to negative
values, which shows that lately the universe has entered into a
phase of accelerated expansion. Finally all the trajectories
asymptotically settle around $q=-1$, which is an expected result
for non-phantom universes. So this model is very efficient in
explaining the evolution of the universe and perfectly
incorporates the observational aspects. We also see that the
dependency of $q$ on the initial conditions becomes less
pronounced in the late universe where all the trajectories nearly
converge.

\begin{figure}\label{fig1}
~~~~~~~~~~~~~~~~~~~~~~~~~~~~~~\includegraphics[height=2.2in,width=3in]{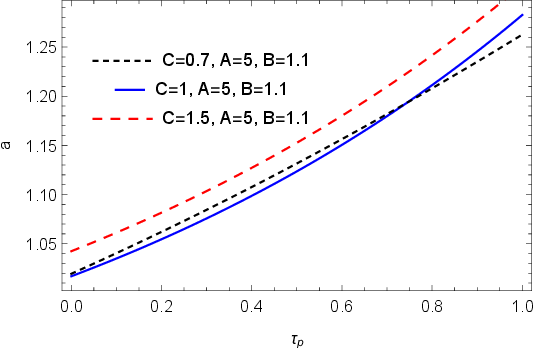}~~~~~\\

~~~~~~~~~~~~~~~~~~~~~~~~~~~~~~~~~~~~~~~~~~~~~~~~~~~~~~~~~~~~Fig.1~~~~~~~~~~~~~~~~~~~~\\

\vspace{1mm} \textit{\textbf{Fig.1} shows the variation of the
scale factor $a$ against the dimensionless time $\tau_p$ for
different initial conditions for model-1. We have considered
$a(0)\approx 1$.}
\end{figure}

\begin{figure}\label{fig2new}
~~~~~~\includegraphics[height=2in,width=2.8in]{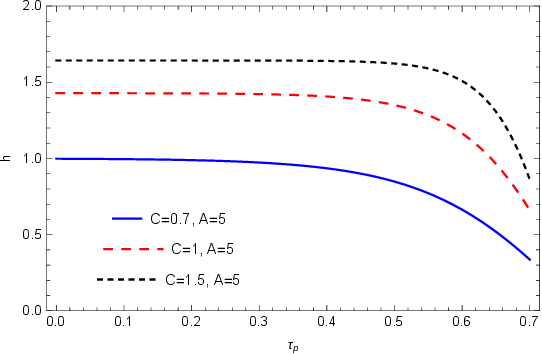}~~~~~~~~~~~~~~\includegraphics[height=2in,width=2.8in]{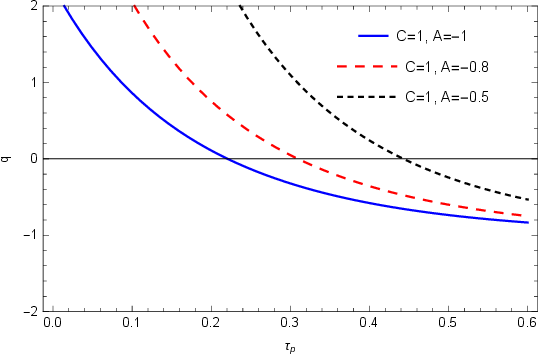}~~~~~\\

~~~~~~~~~~~~~~~~~~~~~~~~~~~~~~~~~Fig.2~~~~~~~~~~~~~~~~~~~~~~~~~~~~~~~~~~~~~~~~~~~~~~~~~~~~~Fig.3~~~~~~~~\\

\vspace{1mm} \textit{\textbf{Fig.2} shows the variation of the
Hubble function $h$ against the dimensionless time $\tau_p$ for
different initial conditions for model-1.}\\

\vspace{1mm} \textit{\textbf{Fig.3} shows the variation of the
deceleration parameter $q$ against the dimensionless time $\tau_p$
for different initial conditions for model-1.}
\end{figure}

For this model the extra force discussed in eqn.(\ref{perfforce})
can be given as (for $n=1$),
\begin{equation}
F^{\mu}_{\mathbf{T}^2}=\frac{\eta\Upsilon^{\mu\nu}}{8\pi\left(\rho+p\right)}\left[D_{\alpha}\left(-
\left(\rho^{2}+4p\rho+3p^{2}\right)u^{\alpha}u_{\nu}\right)-\frac{1}{2}\partial_{\nu}\left(\rho^{2}+3p^2\right)\right]
\end{equation}
The qualitative features of the force component are similar to the
ones discussed for eqn.(\ref{perfforce}). We see that the factor
$\eta$ that appears with the force has a scaling effect on it. In
the STG limit $\eta\rightarrow 0$, which implies
$F^{\mu}_{\mathbf{T}^2}\rightarrow 0$, which is in accordance with
our expectations.

Looking at the model considered here, it is easy to understand
that the model parameter $\eta$ is not dimensionless. Using the
FLRW equations for this model we can perform a dimensional
analysis to find the dimensions of $\eta$. In this work we have
considered Newton's gravitational constant $G$ and the velocity of
light $c$ as unity as can be seen from eqns.(\ref{flrwstan1}) and
(\ref{flrwstan2}). But for dimensional purposes we want to restore
them in Einstein's constant $\kappa$ to get the exact dimensional
results. Generally the constant $\kappa$ in the Einstein's
equations ($G_{\mu\nu}=\kappa T_{\mu\nu}$) is taken as $8\pi
G/c^4$, but this is not unique. Note that in the earlier
computations for $G=c=1$, $\kappa$ has been reduced to $8\pi$. Now
coming to dimensions of the model parameters, we can make use of
the dimensional homogeneity of any physical equation. This implies
that in a meaningful equation describing physics, all the terms of
an equation will have the same dimensions. For model-1, such
equations can be the FLRW equations given by eqns.(\ref{flrwmod1})
and (\ref{flrwmod2}). In eqn.(\ref{flrwmod2}) we will consider
dimensional homogeneity of all the four terms and try to determine
the dimensions of the model parameter $\eta$ from there. We take
the fundamental dimensions as Mass, Length and Time represented by
$[M]$, $[L]$ and $[T]$ respectively. The dimensions of the first
term in the LHS is $[H^2] = [T^{-2}]$. This is because the
dimensions of Hubble parameter H is $[T^{-1}]$ as can be
calculated directly from the Hubble law. The dimensions of the
second term in the LHS is $[\dot{H}] = [T^{-2}]$. Traditionally
the operator $d/dt$ has the inverse dimensions of time, i.e.
$[T^{-1}]$ \cite{dimensions}. So $\dot{H}=dH/dt$ should have the
dimensions $[T^{-1}\times T^{-1}]=[T^{-2}]$. So both terms in the
LHS have the dimensions $[T^{-2}]$. Coming to the second term on
the RHS we see that we can treat it in various ways depending on
what dimensions we choose to consider for the Einstein's constant
$\kappa$ \cite{dimensions}. From ref.\cite{dimensions} we see that
in literature there are various ways to write the Einstein's
constant $\kappa$, which are $\frac{8\pi G}{c^4}$, $\frac{8\pi
G}{c^2}$ and $8\pi G$. Although the first one is the most widely
used form, we have checked that in the present equations the
second form is more suitable. The dimensions of the Newton's
gravitational constant is $[G]=[M^{-1} L^{3} T^{-2}]$ and that of
the velocity of light is $[c]=[LT^{-1}]$. Now using these, we get
the dimensions of $\kappa=8\pi G/c^2$ as $[M^{-1} L]$. The
dimensions of pressure is given by $[p]=[Force]/[area]=[ML^{-1}
T^{-2}]$. Using these we get the dimensions of the second term on
the RHS as $[M^{-1} L \times ML^{-1} T^{-2}]=[T^{-2}]$, which
matches with the dimensions of the terms in the LHS. Finally we
come to the first term in the RHS, which contains the model
parameter $\eta$, whose dimensions we are seeking. We consider $n$
as a dimensionless constant since it appears in the exponent.
Using eqn.(\ref{barotropic}) in (\ref{flrwmod2}) the first term in
the RHS can be written as, $\frac{\eta}{2} p^{2n}
\left(3+\frac{1}{k^2}\right)^n$. Taking the dimensions of the
portion in brackets as $1$ or dimensionless, we are left with
$\frac{\eta}{2} p^{2n}$. Now the overall dimension of this term
must be $[T^{-2}]$ to respect dimensional homogeneity. Therefore
we have $[\eta]\times[p^{2n}]=[T^{-2}]$. From this we see that
$[\eta]=\frac{[T^{-2}]}{[M^{2n} L^{-2n} T^{-4n}]}$. Depending on
the choice of $n$ we can have different dimensional formulas for
$\eta$. For example, if $n=1$, we have
$[\eta]=\frac{[T^{-2}]}{[M^{2}L^{-2} T^{-4}]}=[M^{-2} L^{2}
T^{2}]$.

\subsection{Model: 2}
Here we consider a second model in the product form between the
two scalar invariants $Q$ and $T^{2}$ as below,
\begin{equation}\label{toymode2}
f(Q,T_{\mu\nu}T^{\mu\nu})=f(Q,\textbf{T}^2)=f_{0}Q^{m}
\left(T_{\mu\nu}T^{\mu\nu}\right)^{n}=f_{0}Q^{m}
\left(\textbf{T}^2\right)^{n},~~~~~f_{0}\neq 0
\end{equation}
For $f_{0}\rightarrow 1$, $n\rightarrow 0$ and $m\rightarrow 1$ we
recover the STG limit of the theory. The modified FLRW equations
for this model are given by,
\begin{equation}\label{flrwmod3}
3H^{2}=\left[\frac{2^{4-m}\pi
\rho\left(\rho^{2}+3p^{2}\right)^{1-n}}{f_{0}\left\{\left(6m-6n-3\right)p^{2}
+\left(2m-2n-1\right)\rho^{2}-8np\rho\right\}}\right]^{1/m}
\end{equation}

$$\dot{H}=\frac{1}{2f_{0}m\left(6^{m}H^{2m}+2^{n+1}\times
3^{n}H^{2n}\left(n-1\right)\right)}\times
\left[H\left(\rho^{2}+3p^{2}\right)^{-n-1}\left(3H\left(\rho^{2}+3p^{2}\right)\right.\right.$$

\begin{equation}\label{flrwmod4}
\left.\left.\left(16\pi
p+6^{m}f_{0}H^{2m}\left(2m-1\right)\left(\rho^{2}+3p^{2}\right)^{n}\right)-2^{n+2}\times
3^{n}f_{0}H^{2n}mn\left(\rho^{2}+3p^{2}\right)^{n}\left(\rho\dot{\rho}+3p\dot{p}\right)\right)\right]
\end{equation}
We further simplify these equations by considering $n=m=1$ as
given below,
\begin{equation}\label{flrwmod5}
3H^{2}=-\frac{8\pi\rho}{f_{0}\left(\rho^{2}+8p\rho+3p^{2}\right)}
\end{equation}
\begin{equation}\label{flrwmod6}
2f_{0}\left(\rho^{2}+3p^{2}\right)\dot{H}-3f_{0}\left(\rho^{2}+3p^{2}\right)H^{2}
+4f_{0}\left(\rho\dot{\rho}+3p\dot{p}\right)H=8\pi p
\end{equation}
\\
From the first FLRW equation we directly see that:\\\\
$f_{0}\left(\rho^{2}+8p\rho+3p^{2}\right)<0$ ~~$\Rightarrow$~~
$\left[f_{0}<0\right]$~~ OR~~ $\left[\rho^{2}+8p\rho+3p^{2}<0
\Rightarrow
-\frac{1}{3}\left(4+\sqrt{13}\right)\rho<p<\frac{1}{3}\left(-4+\sqrt{13}\right)\rho\right]$
\\\\
So we readily get a range of the equation of states for the matter
component. This is a very interesting result. We may put the above
equations in the standard form as,
\begin{equation}\label{toy2stan}
3H^{2}=8\pi\rho_{eff}
\end{equation}
\begin{equation}\label{toy2stan}
2\dot{H}+3H^{2}=-8\pi p_{eff}
\end{equation}
where
\begin{equation}\label{toy2effrho}
\rho_{eff}=-\frac{\rho}{f_{0}\left(\rho^{2}+8p\rho+3p^{2}\right)}
\end{equation}
\begin{equation}\label{toy2effp}
p_{eff}=-\frac{1}{8\pi
f_{0}\left(\rho^{2}+3p^{2}\right)}\left[8\pi
p-4f_{0}\left(\rho\dot{\rho}+3p\dot{p}\right)H+6f_{0}\left(\rho^{2}+3p^{2}\right)H^{2}\right]
\end{equation}
From eqn.(\ref{flrwmod5}) we see that $\rho_{c_{3}}=0$ is a
critical point of the model, which corresponds to the empty
universe. Unlike the previous model here we do not get a bouncing
scenario from a non-zero finite energy density. One more
interesting feature of this model is that there are two
singularities corresponding to $p=\frac{1}{3}\left(-4\pm
\sqrt{13}\right)\rho$. Considering barotropic equation of state
given in eqn.(\ref{barotropic}) the above conditions reduce to
$k=\frac{1}{3}\left(-4\pm \sqrt{13}\right)$ for a non-empty
universe. For this model we are able to get an explicit general
solution for any value of $k$. Using eqn.(\ref{barotropic}) in the
eqns.(\ref{flrwmod5}) and (\ref{flrwmod6}) we get,
\begin{equation}\label{soln3}
a(t)=C_{2}\left[\left(k+1\right)^{2}\left(3k-1\right)t+2C_{1}\left(3k^{2}+1\right)\right]^
{-2\left(3k^{2}+1\right)/\left(k+1\right)^{2}\left(3k-1\right)}
\end{equation}
\begin{equation}\label{soln4}
\rho(t)=\frac{6f_{0}\left(3k^{2}+1\right)^{2}\left(3k^{2}+8k+1\right)}{8\pi\left[\left(k+1\right)^{2}\left(3k-1\right)t+2C_{1}\left(3k^{2}+1\right)\right]^{2}}
\end{equation}
where $C_{1}$ and $C_{2}$ are integration constants. It is seen
that the scale factor is undefined at $k=-1$ ($\Lambda$CDM era)
and $k=1/3$ (radiation era). We have generated plots for both
scale factor and dimensionless matter energy density in figures
(4) and (5) respectively. From fig.(4) we see that scale factor
grows with time indicating an expanding universe. We also see that
the expansion rate is dominated in the dark energy phase
$(k=-2/3)$ compared to the other phases, showing signs of
accelerated expansion. Moreover in the late universe the
dependency of $a$ on the initial conditions becomes noteworthy. In
fig.(5) it is evident that the energy density decays with time
which is expected in an expanding universe. Here also the decay
rate is dominated in the exotic phantom phase $(k=-4/3)$ due to
the high rate of expansion of the universe. So the trajectories
are quite satisfactory and comply with the observations. Here the
dependency of $r$ on the initial conditions is quite prominent in
the late times, but not so pronounced in the early universe. We
have also plotted the corresponding trajectories for the standard
$\Lambda$CDM model to get a comparative idea about the parameters
of the model.

\begin{figure}
~~~~~\includegraphics[height=2in,width=2.8in]{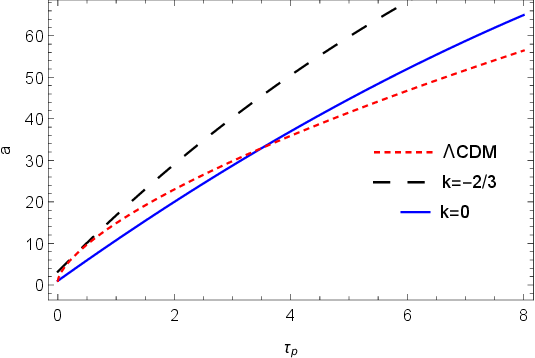}~~~~~~~~~~~~~\includegraphics[height=2in,width=2.8in]{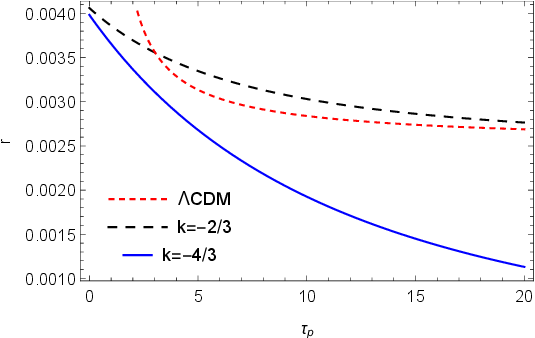}~~~~~\\

~~~~~~~~~~~~~~~~~~~~~~~~~~~~~~Fig.4~~~~~~~~~~~~~~~~~~~~~~~~~~~~~~~~~~~~~~~~~~~~~~~~~~~~~~~~~~~Fig.5~~~~~~~~~~~~~~~~~~~~~~~~~~~~~\\

\vspace{1mm} \textit{\textbf{Fig.4 and 5} shows the variation of
the scale factor $a$ and matter energy density $r$ against the
dimensionless time $\tau_p$ for model-2 for different initial
conditions respectively. For Fig.4 we have taken $C_{1}=5,
C_{2}=-1$. For Fig.5 we have taken $C_{1}=-0.5, f_{0}=0.1,
8\pi=1$. In fig.4 we have considered $a(0)\approx 1$.}
\end{figure}
For the model the continuity equation turns out to be,
\begin{equation}\label{contmod2}
\dot{\rho}+3H\left(\rho+p\right)=3H\left[\left(1+k\right)\rho+\frac{\zeta+f_{0}\pi
\rho^{4}\xi\left(4k\pi-3f_{0}H\xi\rho\right)} {\zeta/\rho+6\pi
f_{0}^{2}\xi^{2}H\rho^{4}}\right]
\end{equation}
where $\zeta=\frac{4}{f_{0}\left(1+8k+3k^{2}\right)}$~~ and~~
$\xi=1+3k^{2}$. The term on the RHS is the non-conservation term.
Setting the non-conservation term equal to zero and solving the
corresponding differential equation for the scale factor we have
the two following evolution of the universe,
\begin{equation}\label{nonconsmod2}
a(t)=C_{3}, ~~a(t)=C_{3}e^{\frac{\xi\sigma\left[-9f_{0}^{2}
\left(\varrho+1\right)\xi^{7}\sigma^{3}-16k\pi^{6}
\left(2C_{1}\xi+\varrho^{2}\left(3k-1\right)t\right)^{8}\right]}
{8\pi^{6}\varrho^{2}\left(6k^{2}+k-1\right)\left[2C_{1}\xi
+\varrho^{2}\left(3k-1\right)t\right]^{9}}}
\end{equation}
where $\sigma=1+8k+3k^{2}$, $\xi=1+3k^{2}$, $\varrho=k+1$ and
$C_{3}$ is the integration constant. The first value suggests that
the scale factor is a constant suggesting a static evolution of
the universe. Since the scale factor does not grow with time the
universe does not expand. This is contrary to our observations. So
we ignore this result. The other expression gives an exponential
type evolution of the universe. It is quite certain that this will
correspond to a de-Sitter like evolution of the universe for
suitable initial conditions. So there exists conditions under
which this model may satisfy the standard energy momentum
conservation relation. The deceleration parameter for this model
is plotted in figure (6). From the figure we see that the there is
a transition of $q$ from positive level to negative level at some
finite value of the dimensionless time $\tau_p$. These values of
$\tau_p$ for each trajectory correspond to the redshift value
$z\approx 0.6$, where the universe enters into the accelerating
phase from a decelerating one. This is cosmologically viable with
the observations. Here the dependency of the deceleration
parameter on the initial conditions grows with time, which is
contrary to the result obtained from the previous model.

\begin{figure}
~~~~~~~~~~~~~~~~~~~~~~~~~~~~~~\includegraphics[height=2.2in,width=3in]{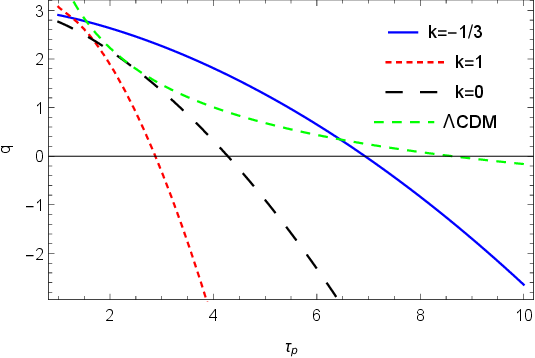}~~~~~\\

~~~~~~~~~~~~~~~~~~~~~~~~~~~~~~~~~~~~~~~~~~~~~~~~~~~~~~~~~~~~Fig.6~~~~~~~~~~~~~~~~~~~~\\

\vspace{1mm} \textit{\textbf{Fig.6} shows the variation of the
deceleration parameter $q$ against the dimensionless time $\tau_p$
for model-2 for different initial conditions. We have taken
$C_{1}=-0.5$, $C_{2}=0.1$, $f_{0}=0.2$, $8\pi=1$.}
\end{figure}

For this model the extra force discussed in eqn.(\ref{perfforce})
can be given as (for $n=1, m=1$),
\begin{equation}
F^{\mu}_{\mathbf{T}^2}=\frac{f_{0}Q\Upsilon^{\mu\nu}}{8\pi\left(\rho+p\right)}\left[D_{\alpha}\left(-
\left(\rho^{2}+4p\rho+3p^{2}\right)u^{\alpha}u_{\nu}\right)-\frac{1}{2}\partial_{\nu}\left(\rho^{2}+3p^2\right)\right]
\end{equation}
Here we see that along with the constant $f_{0}$, the
non-metricity $Q$ also scales the dynamic force component. Since
$Q=6H^{2}$, we see that the expansion factor of the universe has a
direct influence on the force component which was not the case in
the previous model. Moreover we see that with increased expansion
rate the term outside the bracket grows and simultaneously the
term inside the bracket decays due to decreased matter density (as
discussed before). The reverse happens when the expansion rate of
universe decreases. So with one factor growing and the other one
decaying, the force component is likely to evolve into a constant
value with time.

Just like the previous model, here also we are interested in
performing a dimensional analysis to determine the dimensions of
the model parameter $f_{0}$. For this we will use the FLRW
equation given in eqn.(\ref{flrwmod3}). Using
eqn.(\ref{barotropic}) the FLRW equation may be rewritten as,
\begin{equation}\label{dimmod2}
(3H^{2})^{m}f_{0}\rho^{2}\left\{\left(6m-6n-3\right)k^{2}
+\left(2m-2n-1\right)-8nk\right\}=8\pi\times
2^{1-m}\rho(\rho^{2})^{1-n}\left(1+3k^{2}\right)^{1-n}
\end{equation}

Now using the same expression for the Einstein's constant used in
the previous model $\kappa=8\pi G/c^2$ we get the dimensions of
the term in the RHS as $[M^{-1} L] \times
[ML^{-3}]^{3-2n}=[M^{2-2n}L^{6n-8}]$. Here we have used the
dimensions of density as $[\rho]=[mass]/[volume]=[ML^{-3}]$. In
the LHS, considering the dimensions of the term in the bracket as
$1$ (dimensionless) we are left with $(3H^{2})^{m}f_{0}\rho^{2}$
whose dimensions should be $[M^{2-2n}L^{6n-8}]$ so that
dimensional homogeneity is preserved. So we have $[T^{-2m}
]\times[f_{0}]\times[\rho^2]=[M^{2-2n}L^{6n-8}]$, which gives
$[f_{0}]=\frac{[M^{2-2n}L^{6n-8}]}{[T^{-2m}]\times [M^{2}
L^{-6}]}=[M^{-2n} L^{6n-2} T^{2m}]$, which is the dimensional
formula for $f_0$. For a particular case if we consider $m=n=1$,
then the dimensional formula for $f_{0}$ is
$[f_{0}]=[M^{-2}L^{4}T^{2}]$. Finally we would like to state that
the dimensional results derived above (for both the models) is not
unique by any means. It depends on various assumptions and choices
that we have made from time to time.

\section{Energy Conditions}
Energy conditions are tools to establish the positiveness of the
energy-momentum tensor in the presence of matter. These conditions
actually describe the attractive nature of gravity and also take
care of the causal and geodesic structure of the spacetime
\cite{energyc1}. It is known that the energy conditions are
directly linked to GR as they lead to some powerful singularity
theorems \cite{hawk1}. Now the formulation and meaning of energy
conditions in the context of modified gravity theories is an
extremely delicate issue and has its own implications which is
quite contrasting to the implications in GR. Especially the
non-standard (fictitious) fluids related to the additional degrees
of freedom of modified gravity are supposed to produce interesting
results when compiled with the energy conditions, which gives us
some ideas about the non-attractive nature of the gravity leading
to the cosmic acceleration. This is important considering that we
do not yet have a model of cosmology consistent with observations
and free from all the cosmological issues. The prime outcomes are
that matter may manifest further thermodynamical features and
gravity may retain its attractive nature in presence of large
negative pressures. On the other hand, we can have repulsive
gravity for standard matter. The fact that further degrees of
freedom connected with the modified gravity theories, can be dealt
under the banner of effective fluids, does allow us to frame
consistent energy conditions for large classes of theories. From a
cosmological point of view, these considerations are crucial. As
an example, we see that the presence of dark energy can be
considered a direct violation of energy conditions in the standard
sense of GR. However in a generalized approach for modified
gravity theory, there is no such violation, but just a
reinterpretation of the additional degrees of freedom emerging
from the dynamics of the theory \cite{nec1}. So it is clear that
there is a lot to gain in studying energy conditions in modified
gravity. The reader may refer to the Refs.\cite{nec1, nec2} for
further detailed discussions on energy conditions in modified
gravity theories. Moreover energy conditions in the background of
various modified gravity theories may be found in \cite{nec3,
nec4, nec5, nec6, nec7}.

In this section we explore the energy conditions that the
thermodynamic parameters of the $f(Q,\textbf{T}^{2})$ theory need
to satisfy and thus put some constraints on the model parameters.
We will use a perfect fluid matter distribution. It should be
mentioned here that the late cosmic acceleration demands the
violation of the strong energy condition, since it requires
anti-gravitational effect to play its role. The matter component
responsible for this violation may be dark energy. There are
basically four energy conditions that can be derived
from standard general theory of relativity. Considering isotropic cosmology they are:\\\\

(I) Weak Energy Condition (WEC)~$\Rightarrow$~ $\rho_{eff}\geq 0$,~~~~$\rho_{eff}+p_{eff}\geq 0$\\

(II) Null Energy Condition (NEC)~$\Rightarrow$~~~~~$\rho_{eff}+p_{eff}\geq 0$\\

(III) Dominant Energy condition (DEC)~$\Rightarrow$~ $\rho_{eff}\geq 0$,~~~~$\rho_{eff}\geq~ \mid p_{eff} \mid$\\

(IV) Strong Energy condition (SEC)~$\Rightarrow$~ ~~$\rho_{eff}+3p_{eff}\geq 0$\\

Now using eqns.(\ref{rhoeff}) and (\ref{peff}) in the WEC we get
the following inequalities considering $8\pi=1$,
\begin{equation}\label{wec1}
\frac{1}{f_{Q}}\left[\rho+\frac{1}{2}f(Q,\textbf{T}^{2})+f_{\textbf{T}^{2}}\left(\rho^{2}
+4p\rho+3p^{2}\right)\right]\geq 0
\end{equation}
\begin{equation}\label{wec2}
\frac{1}{f_{Q}}\left[3f(Q,\textbf{T}^{2})+4H\left(2\dot{f_{Q}}-9Hf_{Q}\right)+2\left(\rho-2p\right)+2f_{\textbf{T}^{2}}\left(\rho^{2}
+4p\rho+3p^{2}\right)\right]\geq 0
\end{equation}
The expression (\ref{wec2}) is the required condition for the
satisfaction of NEC. The dominant energy
condition~$\rho_{eff}\geq~ \mid p_{eff} \mid$~ may be modified as
$\rho_{eff}\pm p_{eff}\geq 0$. So along with the conditions given
in (\ref{wec1}) and (\ref{wec2}) we have another condition for DEC
given below,
\begin{equation}\label{dec1}
\frac{1}{f_{Q}}\left[f(Q,\textbf{T}^{2})+4H\left(2\dot{f_{Q}}-9Hf_{Q}\right)
-2\left(\rho+2p\right)-2f_{\textbf{T}^{2}}\left(\rho^{2}
+4p\rho+3p^{2}\right)\right]\leq 0
\end{equation}
Finally from the SEC we get,
\begin{equation}\label{sec1}
\frac{1}{f_{Q}}\left[7f(Q,\textbf{T}^{2})+12H\left(2\dot{f_{Q}}-9Hf_{Q}\right)
+2\left(\rho-6p\right)+2f_{\textbf{T}^{2}}\left(\rho^{2}+4p\rho+3p^{2}\right)\right]\geq
0
\end{equation}
So all the energy conditions finally yield four inequalities given
by (\ref{wec1}), (\ref{wec2}), (\ref{dec1}) and (\ref{sec1}),
which can be used to constrain the theory. Now we may use our
toy-models discussed above to check how viable and effective these
energy conditions are in constraining cosmological models. We
discuss them one by one below.

\subsection{Model: 1}
Here we will use the model given in eqn.(\ref{toymodel}) which has
two free parameters $n$ and $\eta$. We will also consider the
barotropic equation of state $p=k\rho$. Using the above relations
we give the energy conditions for this model as,\\

$$\bullet~~ \textbf{WEC:}~~3H_{0}^{2}+\rho_{0}+\frac{1}{2}\left(1+3k^{2}\right)^{n-1}
\left[1+2n+k\left(3k+8n+6kn\right)\right]\eta\rho_{0}^{2n}\geq 0$$
~~~~~~~~~~and
\begin{equation}\label{wecm1}
-18H_{0}^{2}+2\left(1-2k\right)\rho_{0}+\left(1+3k^{2}\right)^{n-1}\left[3+2n
+k\left(9k+8n+6kn\right)\right]\eta\rho_{0}^{2n}\geq 0
\end{equation}

\begin{equation}\label{necm1}
\bullet~~\textbf{NEC:}~~-18H_{0}^{2}+2\left(1-2k\right)\rho_{0}+\left(1+3k^{2}\right)^{n-1}\left[3+2n
+k\left(9k+8n+6kn\right)\right]\eta\rho_{0}^{2n}\geq 0
\end{equation}

$$\bullet~~\textbf{DEC:}~~3H_{0}^{2}+\rho_{0}+\frac{1}{2}\left(1+3k^{2}\right)^{n-1}
\left[1+2n+k\left(3k+8n+6kn\right)\right]\eta\rho_{0}^{2n}\geq
0$$,

$$-18H_{0}^{2}+2\left(1-2k\right)\rho_{0}+\left(1+3k^{2}\right)^{n-1}\left[3+2n
+k\left(9k+8n+6kn\right)\right]\eta\rho_{0}^{2n}\geq 0$$

and

\begin{equation}\label{decm1}
\left(1+3k^{2}\right)^{n-1}\left[2n+8kn+3k^{2}\left(2n-1\right)-1\right]\eta\rho_{0}^{2n}
+2\left(15H_{0}^{2}+\rho_{0}+2k\rho_{0}\right)\geq 0
\end{equation}

\begin{equation}\label{necm1}
\bullet~~\textbf{SEC:}~~2\left(1-6k\right)\rho_{0}+\left(1+3k^{2}\right)^{n-1}\left[7+2n
+k\left(21k+8n+6kn\right)\right]\eta\rho_{0}^{2n}-66H_{0}^{2}\geq
0
\end{equation}
Since we are interested in constraining the models we have used
the present values of the Hubble parameter $H_{0}$ and matter
energy density $\rho_{0}$ in the above energy conditions. The
present value of the Hubble parameter is estimated to be
$H_{0}=67.9~ km sec^{-1}Mpc^{-1}$ \cite{planck, capo2} and that of
$\rho_{0}$ is $\rho_{0}=9.9\times 10^{-30} ~gm ~cm^{-3}$
\cite{nasa}. Now we may consider various cosmological era for
matter by changing the value of $k$ such as $k=1/3, 0, -1/3, -1$.
After putting all these values we will get inequalities connecting
only $\eta$ and $n$, from where it will be straightforward to put
constraints on these two model parameters.

\subsubsection{$k=1/3$ (Radiation)}
From the WEC conditions we found that $-0.2 \leq n \leq -0.1$ and
$\eta>0$. From further analysis it is evident that the above range
of the parameters also satisfy the other conditions NEC, DEC and
SEC.

\subsubsection{$k=0$ (Dust)}
For this cosmological era it is found that the parameter range
obtained from the WEC conditions are $-0.5 \leq n \leq -0.1$ and
$\eta>0$. Using the other conditions the range was reduced to
$-0.5 \leq n < -0.1$ and $\eta>0$.

\subsubsection{$k<-1/3$ (Quintessence)}
The constraints on the parameter space for this era are found as
$n \leq -0.1$ and $\eta>0$ from the WEC conditions. The other
conditions comply with this range with the exception of the SEC
condition. This is consistent with the accelerated expansion of
the universe.

\subsection{Model: 2}
In this model we have three free parameters $f_{0}$, $m$ and $n$.
Since for this model we have $\dot{H}$ present in $\dot{f_{Q}}$
terms, we have used the eqn.(\ref{decp}) to define
$\dot{H}=-H^{2}\left(1+q\right)$. Below we present the energy conditions for this model.\\

$\bullet$~~\textbf{WEC:}~~\\
$$\frac{8kn+\left(3k^{2}+1\right)\left(1+2n\right)}
{m}+\frac{2\times\left(6H_{0}^{2}\right)^{-m}\rho_{0}^{1-2n}
}{f_{0}m\left(1+3k^{2}\right)^{n-1}} \geq 0,$$ ~~~~~~~~~~and

\begin{equation}\label{wecm2}
\frac{1}{m}\left[3+2\varrho\left(3\varrho-2\right)
\xi^{-1}n+2f_{0}^{-1}H_{0}^{-2m}\xi^{-n}\left(3-2\varrho\right)6^{-m}\rho_{0}^{1-2n}+\frac{2}{3}m
\left\{2(6H_{0}^{2})^{(n-m)}
\left(2\left(1-n\right)q_{0}+\frac{\varsigma}{\vartheta}\right)-9\right\}\right]\geq
0
\end{equation}

$\bullet$~~\textbf{NEC:}~~\\
\begin{equation}\label{necm2}
\frac{1}{m}\left[3+2\varrho\left(3\varrho-2\right)
\xi^{-1}n+2f_{0}^{-1}H_{0}^{-2m}\xi^{-n}\left(3-2\varrho\right)6^{-m}\rho_{0}^{1-2n}+\frac{2}{3}m
\left\{2(6H_{0}^{2})^{(n-m)}
\left(2\left(1-n\right)q_{0}+\frac{\varsigma}{\vartheta}\right)-9\right\}\right]\geq
0
\end{equation}

$\bullet$~~\textbf{DEC:}~~\\
$$\frac{8kn+\left(3k^{2}+1\right)\left(1+2n\right)}
{m}+\frac{2\times\left(6H_{0}^{2}\right)^{-m}\rho_{0}^{1-2n}
}{f_{0}m\left(1+3k^{2}\right)^{n-1}} \geq 0,$$

\begin{equation}\label{decm2}
\frac{1}{m}\left[7+2\varrho\left(3\varrho-2\right)
\xi^{-1}n+2f_{0}^{-1}H_{0}^{-2m}\xi^{-n}\left(3-2\varrho\right)6^{-m}\rho_{0}^{1-2n}+\frac{2}{3}m
\left\{2(6H_{0}^{2})^{(n-m)}
\left(2\left(1-n\right)q_{0}+\frac{\varsigma}{\vartheta}\right)-9\right\}\right]\geq
0,
\end{equation}
and

\begin{equation}\label{decm2}
\frac{1}{m}\left[1-2\varrho\left(3\varrho-2\right)
\xi^{-1}n-2f_{0}^{-1}H_{0}^{-2m}\xi^{-n}\left(3-2\varrho\right)6^{-m}\rho_{0}^{1-2n}+\frac{2}{3}m
\left\{2(6H_{0}^{2})^{(n-m)}
\left(2\left(1-n\right)q_{0}+\frac{\varsigma}{\vartheta}\right)-9\right\}\right]\leq
0
\end{equation}

$\bullet$~~\textbf{SEC:}~~\\
\begin{equation}\label{secm2}
\frac{1}{m}\left[3+2\varrho\left(3\varrho-2\right)
\xi^{-1}n+2f_{0}^{-1}H_{0}^{-2m}\xi^{-n}\left(7-6\varrho\right)6^{-m}\rho_{0}^{1-2n}+2m
\left\{2(6H_{0}^{2})^{(n-m)}
\left(2\left(1-n\right)q_{0}+\frac{\varsigma}{\vartheta}\right)-9\right\}\right]\geq
0
\end{equation}
In the above expressions $\sigma$, $\xi$ and $\varrho$ have been
defined earlier just after eqn.(\ref{nonconsmod2}) and the
expressions for $\varsigma$ and $\vartheta$ are given below,

$$\varsigma=(\xi\rho_{0}^{2})^{-n}\left[-2f_{0}\left((6H_{0}^{2})^{m}
+4(6H_{0}^{2})^{n}\left(n-1\right)\right)\left(1-3\left(m+1\right)n+2n^{2}\right)
(\xi\rho_{0}^{2})^{n}\chi\right.$$

\begin{equation}
\left.+6nH_{0}^{2}\left(-6k\rho_{0}-f_{0}\left(3(6H_{0}^{2})^{m}
\left(3m-1\right)+2(6H_{0}^{n})^{n}m\left(n-1\right)\right)(\xi\rho_{0}^{2})^{n}\right)\right]
\end{equation}

\begin{equation}
\vartheta=f_{0}\left[-2mn(6H_{0}^{2})^{1+n}+\left((6H_{0}^{2})^{m}
+2(6H_{0}^{2})^{n}\left(n-1\right)\right)2\left(2n-1\right)\chi\right]
\end{equation}

where
$$\chi=\left(\frac{6^{-m}(\xi\rho_{0}^{2})^{n-1}}{f_{0}\rho_{0}
\left(\xi\left(2m-1\right)-2\varrho\left(3\varrho-2\right)n\right)}\right)^{1/m}$$

Just like the previous model, in addition to the current values
$H_{0}$, $\rho_{0}$ we also have to use~ $q_{0}=-0.503$~
\cite{planck, capo2} for this case, to put constraints on the
parameters $f_{0}$, $n$ and $m$ from the above inequalities (which
is pretty straightforward). So the above energy conditions can be
used as relations to constrain the free parameters of the theory
as done for model-1. Since the conditions for this model are quite
complex, we have managed to obtain some general constraints on the
model parameters. The results are presented below in a tabular
form.

\vspace{4mm}
\begin{center}
\begin{tabular}{|l||l|l|}
\hline

~$k$~&~~~~~Range of $f_{0}$~&~~~~~Constraint on model parameters~\\
\hline
            &           &
\\

~$1/3$~&~$\bullet$ ~~For $f_{0}>0$ ~&~~~~$(i)~ n\geq -0.2,~ m>0~~~~  (ii)~ n\leq -0.3,~m<0$~\\
~$~$~&~$~$~&~$~$~\\
~&~$\bullet$ ~~For $f_{0}<0$~&~~~~$(i)~ n<-0.2,~ m<0 ~~~~ (ii)~ n\geq -0.2,~ m>0$~ \\
~$~$~&~$~$~&~$~$~\\

\hline
            &          &
\\

~$0$~&~$\bullet$ ~~For $f_{0}>0 $~&~~~~$(i)~ n\geq -0.4,~ m>0$~\\
~$~$~&~$~$~&~$~$~\\
~&~$\bullet$ ~~For $f_{0}<0$~&~~~~$(i)~ n<0,~ m<0 ~~~~ (ii)~ n>0,~ m \leq -0.1$~\\
~$~$~&~$~$~&~$~$~\\

\hline
            &           &

\\

$-2/3$~&~$\bullet$ ~~For $f_{0}>0$~&~~~~~$m>0,$ ~~for all values of $n$~\\
~$~$~&~$~$~&~$~$~\\
~&~$\bullet$ ~~For $f_{0}<0$~&~~~~~$n<0,~ m>0$~\\
~$~$~&~$~$~&~$~$~\\
\hline

\end{tabular}
\end{center}
\vspace{3mm} {\bf Table 1:} Constraints on model parameters (model
2) for different values of EoS parameter $k$ from the energy
conditions. \vspace{6mm}

It should be noted that the parameter values used in the
discussion in section 4.1 violate the values obtained from the
energy conditions in section 5.1. Energy conditions are basically
some mathematically imposed boundary conditions that help us to
deduce very powerful and general results regarding the behaviour
of strong gravitational fields and cosmological geometries
\cite{ec1}. But of late these conditions have started to look far
less secure than once they seemed to be. There can be various
reasons behind this. There are subtle quantum effects which are
responsible for the violation of the energy conditions. There are
also certain classical systems that violate all the energy
conditions \cite{ec1, ec2, ec3}. This directly reflects on the
nature of the matter content of the universe and opens up various
exotic possibilities such as traversable wormholes, warp drives,
time machines, etc. \cite{ec1}. Over the years energy conditions
like the Trace energy condition (TEC) have totally lost their
significance and have now been abandoned. With the discovery of
cosmic acceleration and consequent arrival of the concept of dark
energy, SEC and NEC have almost been abandoned. So the place of
energy conditions in GR and Cosmology needs a radical
reassessment.

From the above discussion it is clear that in a late accelerating
universe filled with dark energy it is expected that SEC and NEC
will be violated. It is known that the early universe was
dominated by quantum effects. For the theory we are discussing in
this paper, there is clear evidence of quantum gravitational
effects in the early universe. These quantum fluctuations are
responsible for the violation of the energy conditions in the
early universe. Moreover for the inflationary epoch, some of the
energy conditions are readily violated. Coming to the period
between the inflation and the late cosmic acceleration (we call
middle phase), it can be argued that due to a strong quantum
gravitational effect in the early epoch, there are some
reminiscent effects in subsequent eras. This imprint of quantum
effects in the system does not allow the energy conditions to
hold. Moreover it has been already discussed that the violation of
energy conditions is also true for certain classical systems as
well \cite{ec1, ec2, ec3}. So even if there is no direct quantum
dominance or dark energy dominance during the middle phase, it is
not very strange for the system to violate the energy conditions.
For DEC and WEC, $\rho_{eff}\geq 0$ will always hold. But the
trouble is with the conditions $\rho_{eff}+p_{eff}\geq 0$ and
$\rho_{eff}\geq |p_{eff}|$. For sufficient negative pressure these
two conditions are violated leading to the overall violation of
WEC and DEC. Our model seems to violate all the four energy
conditions and this is not very unexpected from the above
discussion.

\section{Discussion \& Conclusion}
In the present work we have proposed yet another extension of the
symmetric teleparallel gravity by generalizing the gravity
Lagrangian with an arbitrary function $f(Q,T_{\mu\nu}T^{\mu\nu})$.
The field equations were derived in a metric-affine formalism. The
correction terms introduced by the modified gravity were noted. As
expected for any gravity theory involving non-minimal coupling
between geometry and matter sectors, the covariant divergence of
the energy-momentum tensor was non-zero thus implying the
non-conservation of the same. The non-conservation term was
derived using the field equations. The momentum conservation
equation showed the presence of correction terms implying extra
force on the massive particles thus making the motion
non-geodesic. The field equations were further simplified by
considering perfect fluid as the matter component. Using these
field equations we resorted to study the cosmological evolution of
the theory. The FLRW equations for a flat homogeneous and
isotropic spacetime were derived. It was noted that there were two
additional modification terms introduced in the equations in
contrast to those of standard GR. The two additional terms were of
the form $f/2$ and
$f_{\textbf{T}^{2}}\left(\rho^{2}+4p\rho+3p^{2}\right)$. The first
one came from the coupling between the geometric and matter sector
and the second one is completely a source term. These higher order
terms dominate in the early universe and gradually fade away at
late times giving the effects of the standard FLRW universe. These
corrections are totally intrinsic and uniquely describe the
modified gravity. Expressions for some cosmological parameters
like the equation of state and deceleration parameter were
derived. We investigated the vacuum solution of the theory and saw
that EMSSTG admits a de-Sitter like solution in its framework. One
of the crucial aspect of the theory is the non-conservation of the
energy-momentum tensor. Moreover in the momentum conservation
equation, an extra force appears which results in non-geodesic
motion of massive particles.

To get more insights into the cosmological framework of the theory
we studied two specific toy-models models $Q+\eta
\left(\textbf{T}^2\right)^{n}$ and $f_{0}Q^{m}
\left(\textbf{T}^2\right)^{n}$. We saw that both the models had
STG as a limiting case and hence we can recover the parent theory
from the equations. After deriving the FLRW equations for the
first model we saw that the equations had a flavour of the quantum
gravity effects of the loop quantum gravity. So solutions for
bouncing cosmology was investigated and it was found that the
model indeed supported a cosmological bounce at a finite time,
thus avoiding the singularity. Various constraints were imposed on
the model from these relations. Although we did not get a general
solution of the model, but for $k=-1$ we obtained a solution which
resembled the de-Sitter solution. Then we derived the continuity
equation for the model and studied the non-conservation term.
Using it we were able to trace the evolution of the scale factor
for which the non-conservation term will vanish. For this model we
plotted the scale factor, the Hubble function and the deceleration
parameter to check the viability of the model. Similar studies
were undertaken for the second model. In this case we were
fortunate enough to get a general solution of the FLRW equations.
The obtained scale factor and the matter energy density were
plotted and compared to those of the standard $\Lambda$CDM model.
We also obtained the plots for the deceleration parameter for this
model in a comparative scenario with the $\Lambda$CDM model. The
transition from a decelerating to an accelerating universe was
clearly evident and the deviation of the trajectories from those
of the $\Lambda$CDM model was also noted. All the plots are
generated using dimensionless parameters. For both the models a
detailed dimensional analysis is performed to determine the
dimensions of the model parameters.

It must be stated here that for our theory, the presence of extra
force and the corresponding non-geodesic motion of the test
particle, implies the violation of the equivalence principle (EP).
There are weak and strong forms of the EP. Even it is accepted by
many authors that although most of the metric theories of gravity
satisfy the weak form of the principle, GR is the only gravity
theory in four dimensions that fully incorporates the strong
equivalence principle (SEP) \cite{ep1, ep2}. So if we are
searching for concepts beyond GR, a promising avenue will be to
look for occasions of the violation of the SEP. To complement this
we would like to mention that gravity's rainbow \cite{ep3}, which
is an extension of the doubly special relativity \cite{ep4} to
incorporate curvature, is a quantum theory of gravity, where there
is a direct violation of the EP. In this theory the path followed
by a particle in a gravitational field depends on the energy
content of the body and hence there is a modification to the
standard EP. Since our theory also has flavours of quantum gravity
it is quite expected from analogy that there should be some
confrontation with the EP and possible violation. Coming to the
tests, there has been no universal acceptance backed by
experimental observations of the EP till date. This is evident
from the fact that people are continuing to test the principle
till date and trying to find ways to prove its validity or
disprove it \cite{ep5, ep6}. This shows that, may be the tests are
not yet well framed and self consistent or probably our
instruments are not yet advanced enough to test the theory, but
obviously progress is being made. Also we just cannot rule out the
fact that the EP may not be true. We don't know for the time being
and neither can we claim anything. That is why these alternative
theories with non-minimal matter coupling have gained importance
over the past decade. We need a competing concept to challenge the
existing one (at least in the absence of a proof). So for the time
being there is place for counter concepts and these are not rare
in literature. Our model is nothing new, but just another elegant
example of it having very important and impressive properties like
quantum gravity. We can also think of modifications to the EP
(like gravity's rainbow) consistent with these non minimal matter
coupling theories and try to test them. It has also been reported,
from the data of the Abell Cluster A586, that interaction of dark
matter and dark energy does imply the violation of the EP
\cite{ep7}. Thus there is a realistic possibility and
justification of studying and testing these models with
non-minimal matter couplings in the context of the violation of
the EP.

Finally we explored the energy conditions in the background of the
theory. The basic energy conditions WEC, NEC, DEC and SEC were
derived for the theory and also for the two specific toy-models.
In the first model we had two free parameters $n$ and $\eta$ after
we fixed $H$ and $\rho$ with their current values from the
observations. Similarly in the second model we had three free
parameters $f_{0}$, $m$ and $n$ after the fixation from the
observations. Basically four constraints were obtained for each
model using the energy conditions which are sufficient to put
bounds on the free parameters of both the models. Thus the models
were well-constrained by the energy conditions. It was also found
that the energy conditions are violated in the cosmological
discussion of the models, which is quite expected given the
quantum gravity effects and the exotic nature of the theory.
Finally from the study we conclude that the theory is perfectly
suitable to describe the cosmological dynamics of both the early
and the late universe without resorting to dark energy. There is
scope for further development of the theory, which will be
undertaken in future projects. The standout feature of EMSSTG is
that the field equations contain terms which arise from the
quantum gravity effects and thus are responsible for the avoidance
of the singularity. So this theory is a singularity free
cosmologically viable theory. Moreover the non-linear density
terms in the equations dominate in the early universe and
gradually fade away at later times. So the quantum effects of
modified gravity is predominantly felt in the early universe and
it eases out to give the standard FLRW effects at late times.

\section*{Acknowledgments}

The author acknowledges the Inter University Centre for Astronomy
and Astrophysics (IUCAA), Pune, India for granting visiting
associateship. The author also thanks the anonymous referee for
his/her invaluable comments that helped to improve the quality of
the manuscript.



\begin{thebibliography}{99}

\bibitem{gr} A. Einstein :- {\it Annalen der Physik} {\bf 354}, 769 (1916).
\bibitem{eddo} D. F. Watson, E. A. Stanley, Davidson C.:- {\it Philosophical Transactions of the Royal Society of London: Series A} {\bf 220} 291 (1920)
\bibitem{redshift1} L. D. Landau, E. M. Lifshitz, The Classical Theory of Fields (Pergamon Press, Oxford, 1970)
\bibitem{gw1} B. P. Abbott et al., LIGO Scientific Collaboration and Virgo Collaboration:- {\it Phys. Rev. Lett.} {\bf 116}, 061102 (2016).
\bibitem{ehtc} The Event Horizon Telescope Collaboration :- {\it ApJL} {\bf 875} L1 (2019)
\bibitem{tests1} C.M. Will :- {\it Living Rev. Relat.} {\bf 17}, 4 (2014)
\bibitem{tests2} M. Ishak :- {\it Living Rev. Relat.} {\bf 22}, 1 (2019).
\bibitem{acc1} S. Perlmutter et. al. :- {\it Astrophys. J.} {\bf 517} 565 (1999).
\bibitem{acc2} A. G. Riess et al. :- {\it Astron. J.} {\bf 116} 1009 (1998).
\bibitem{de1} P. Brax :- {\it Rep. Prog. Phys.} {\bf 81} 016902 (2018)
\bibitem{mod1} S. Nojiri, S. D. Odintsov, V. K. Oikonomou :- {\it Phys. Rep.} {\bf 692} 1 (2017).
\bibitem{mod2} S. Nojiri and S. D. Odintsov :- {\it Int. J. Geom. Methods Mod. Phys.} {\bf 04} 115 (2007).
\bibitem{mod3} S. Capozziello, R. D'Agostino, O. Luongo:- {\it Int. J. Mod. Phys. D} {\bf 28} 1930016 (2019).
\bibitem{fr1} H. A. Buchdahl : {\it Month. Not. R. Astron. Soc.} {\bf 150}, 1 (1970)
\bibitem{fr2} L. Amendola, D. Polarski, S. Tsujikawa :- {\it Phys. Rev. Lett.} {\bf 98} 131302 (2007).
\bibitem{fr3} T. P. Sotiriou :- {\it Classical Quantum Gravity} {\bf 23} 5117 (2006).
\bibitem{fr4} Y.-S. Song, W. Hu, I. Sawicki :- {\it Phys. Rev. D} {\bf 75} 044004 (2007).
\bibitem{fr5} S. Nojiri, S. D. Odintsov :- {\it Phys. Rev. D} {\bf 74} 086005 (2006).
\bibitem{fr6} P. Rudra :- {\it Nucl. Phys. B} {\bf 956} 115014 (2020)
\bibitem{fr7} P. Rudra :- {Commun. Theor. Phys.} {\bf 66} 149 (2016)
\bibitem{frrev1} T. P. Sotiriou, V. Faraoni :- {\it Rev. Mod. Phys.} {\bf 82} 451 (2010).
\bibitem{frrev2} A. De Felice, S. Tsujikawa :- {\it Living Rev. Relativity} {\bf 13} 3 (2010).
\bibitem{frlm1} T. Harko, F. S. N. Lobo :- {\it Eur. Phys. J. C.} {\bf 70} 373 (2010).
\bibitem{frlm2} R. Ribeiro, J. Páramos :- {\it Phys. Rev. D} {\bf 90} 124065 (2014).
\bibitem{frlm3} R. P. L. Azevedo, J. Páramos :- {\it Phys. Rev. D} {\bf 94} 064036 (2016).
\bibitem{frlm4} B. Pourhassan, P. Rudra :- {\it Phys. Rev. D} {\bf 101} 084057 (2020).
\bibitem{harko1} T. Harko, F. S. N. Lobo, S. Nojiri, S. D. Odintsov :- {\it Phys. Rev. D.} {\bf 84} 024020 (2011).
\bibitem{frt1} M. Sharif, M. Zubair :- {\it JCAP} {\bf 03} 028 (2012).
\bibitem{frt2} E. H. Baffou, M. J. S. Houndjo, M. E. Rodrigues, A. V. Kpadonou, J. Tossa :- {\it Phys. Rev. D} {\bf 92} 8, 084043 (2015).
\bibitem{frt3} P. Rudra :- {\it Eur. Phys. J. Plus} {\bf 130}  4, 66 (2015).
\bibitem{frt4} H. Shabani, M. Farhoudi :- {\t Phys. Rev. D} {\bf 88} 044048 (2013).
\bibitem{frt5} F. G. Alvarenga, A. de la Cruz-Dombriz, M. J. S. Houndjo, M. E. Rodrigues, D. Sáez-Gómez :- {\it Phys. Rev. D} {\bf 87}  10, 103526 (2013).
\bibitem{frt6} P. Rudra, K. Giri :- {\it Nucl. Phys. B} {\bf 967} 115428 (2021)
\bibitem{frt7} P. Rudra :- arxiv: 2006.00228 [gr-qc]
\bibitem{emsgorg1} N. Katirci, M. Kavuk, :- {\it Eur. Phys. J. Plus} {\bf 129}, 163(2014)
\bibitem{emsgorg} M. Roshan, F. Shojai :- {\it Phys. Rev. D.} {\bf 94} 044002 (2016).
\bibitem{lqg1} A. Ashtekar, T. Pawlowski, P. Singh :- {\it Phys. Rev. D} {\bf 74}, 084003 (2006).
\bibitem{board1} C. V. R. Board, J. D. Barrow :- {\it Phys. Rev. D} {\bf 96}, 123517 (2017).
\bibitem{akarsu2} O. Akarsu, N. Katirci, S. Kumar, R. C. Nunes, M. Sami :- {\it Phys. Rev. D} {\bf 98} 6 (2018)\\
\bibitem{rudra1} S. Bahamonde, M. Marciu, P. Rudra :- {\it Phys. Rev. D} {\bf 100}, 083511  (2019).
\bibitem{rudra2} C. Ranjit, P. Rudra, S. Kundu :- {\it Annals Phys.} {\bf 428} 168432 (2021).
\bibitem{rudra3} P. Rudra, B. Pourhassan :- {\it Phys. Dark Univ.} {\bf 33} 100849 (2021).
\bibitem{akarsu3} O. Akarsu, N. Katirci, S. Kumar :- {\it Phys. Rev. D} {\bf 97} 2 (2018)\\
\bibitem{akarsu1} O. Akarsu, J. D. Barrow , S. Cikintoglu, K. Y. Eksi, N. Katirci :- {\it Phys. Rev. D} {\bf 97} 12 (2018)\\
\bibitem{moraes1} P. H. R. S. Moraes , P. K. Sahoo :-  {\it Phys. Rev. D} {\bf 97} 2 (2018)\\
\bibitem{nari1} N. Nari, M. Roshan :- {\it Phys. Rev. D} {\bf 98} 2 (2018)\\
\bibitem{keskin1} A. Keskin :-  {\it AIP Conf. Proc.} {\bf 2042} 1 (2018)\\
\bibitem{akarsu4} O. Akarsu, J. D. Barrow, C. V. R. Board, N. M. Uzun, J. A. Vazquez :- {\it Eur. Phys. J. C.} {\bf 79} 10, 846 (2019)\\
\bibitem{riemanng} B. Riemann, Habilitationsschrift, 1854. Abhandlungen der Königlichen Gesellschaft der Wissenschaften zu Göttingen {\bf 13}, 1 (1867)
\bibitem{weyl} H. Weyl :- {\it Sitzungsber. Preuss. Akad. Wiss.} {\bf 465}, 1 (1918).
\bibitem{cartan1} E. Cartan, C. R. Acad :- {\it Sci.} {\bf 174}, 593 (1922).
\bibitem{cartan2} E. Cartan :- {\it Ann. Ec. Norm.} {\bf 40}, 325 (1923).
\bibitem{cartan3} E. Cartan :- {\it Ann. Ec. Norm.} {\bf 41}, 1 (1924).
\bibitem{cartan4} E. Cartan :- {\it Ann. Ec. Norm.} {\bf 42}, 17 (1925).
\bibitem{ecrev} F. W. Hehl, P. von derHeyde, G. D. Kerlick, J. M. Nester : {\it Rev. Mod. Phys.} {\bf 48}, 393 (1976).
\bibitem{wc1} D. Puetzfeld, R. Tresguerres :- {\it Class. Quant. Gravity} {\bf 18}, 677 (2001).
\bibitem{wc2} D. Putzfeld :- {\it Class. Quant. Gravity} {\bf 19}, 4463 (2002).
\bibitem{wc3} D. Puetzfeld :- {\it Class. Quant. Gravity} {\bf 19}, 3263 (2002).
\bibitem{wc4} T. Y. Moon, P. Oh, J.S. Sohn :- {\it JCAP} {\bf 11}, 005 (2010).
\bibitem{ecwcrev} M. Novello, S.E. Perez Bergliaffa :- {\it Phys. Rep.} {\bf 463}, 127 (2008).
\bibitem{weit} R. Weitzenböck :- {\it Invariantentheorie} (Noordhoff, Groningen, 1923)
\bibitem{tegr1} A. Einstein :- Preussische Akademie der Wissenschaften, Phys.-math. Klasse, Sitzungsberichte 1928, 217 (1928).
\bibitem{tegr2} C. Moller :- {\it Mat. Fys. Skr. Dan. Vid. Selsk.} {\bf 1}, 10 (1961)
\bibitem{tegr3} C. Pellegrini, J. Plebanski :- {\it Mat. Fys. Skr. Dan. Vid. Selsk.} {\bf 2}, 4 (1963)
\bibitem{tegr4} K. Hayashi, T. Shirafuji :- {\it Phys. Rev. D} {\bf 19}, 3524 (1979)
\bibitem{ft1} Y-F. Cai, S. Capozziello, M. D. Laurentis, E. N. Saridakis :- {\it Rep. Prog. Phys.} {\bf 79} 106901 (2016)
\bibitem{ft2} M. Krssak, R. J. van den Hoogen, J. G. Pereira, C. G. Bohmer, A. A. Coley :- {\it Class. Quantum Grav.} {\bf 36} 183001 (2019)
\bibitem{fta1} R. Ferraro, F. Fiorini :- {\it Phys. Rev. D} {\bf 75}, 084031 (2007)
\bibitem{fta2} R. Ferraro, F. Fiorini :- {\it Phys. Rev. D} {\bf 78}, 124019 (2008)
\bibitem{fta3} G. R. Bengochea, R. Ferraro :- {\it Phys. Rev. D} {\bf 79}, 124019 (2009)
\bibitem{fta4} E. V. Linder :- {\it Phys. Rev. D} {\bf 81}, 127301 (2010)
\bibitem{fta5} T. Harko, F. S. N. Lobo, G. Otalora, E. N. Saridakis :- {\it Phys. Rev. D} {\bf 89}, 124036 (2014)
\bibitem{fta6} S. Bahamonde, C. G. Boehmer, M. Krssak :- {\it Phys. Lett. B} {\bf 775}, 37 (2017)
\bibitem{fta7} M. Jamil, D. Momeni, R. Myrzakulov, P. Rudra :- {\it J. Phys. Soc. Jap.} {\bf 81} 114004 (2012)
\bibitem{fta8} P. Rudra :- {\it Astrophys. Space Sci.} {\bf 357} 135 (2015)
\bibitem{fta9} C. Ranjit, P. Rudra :- {Int. J. Mod. Phys. D} {\bf 25} 1650008 (2016)
\bibitem{revsb} S. Bahamonde, K. F. Dialektopoulos, C. Escamilla-Rivera et al. :- arxiv: 2106.13793 [gr-qc] (2021)
\bibitem{wcw1} Z. Haghani, T. Harko, H. R. Sepangi, S. Shahidi :- {\it JCAP} {\bf 10}, 061 (2012)
\bibitem{wcw2} Z. Haghani, T. Harko, H. R. Sepangi, S. Shahidi :- {\it Phys. Rev. D} {\bf 88}, 044024 (2013)
\bibitem{st1} J. M. Nester, H.-J. Yo :- {\it Chin. J. Phys.} {\bf 37}, 113 (1999)
\bibitem{cgr1} J. Beltran Jimenez, L. Heisenberg, T. Koivisto :- {\it Phys. Rev.} {\bf 98}, 044048 (2018)
\bibitem{fq1} J. Lu, X. Zhao, G. Chee :- {\it Eur. Phys. J. C} {\bf 79}, 530 (2019)
\bibitem{fq2} R. Lazkoz, F. S. N. Lobo, M. Ortiz-Baño, V. Salzano :- {\it Phys. Rev. D} {\bf 100} 104027 (2019)
\bibitem{stegr2} M. Adak, M. Kalay, O. Sert :- {\it Int. J. Mod. Phys. D} {\bf 15}, 619 (2006).
\bibitem{stegr3} M. Adak :- {\it Turk. J. Phys.} {\bf 30}, 379 (2006).
\bibitem{stegr4} M. Adak, Ö. Sert, M. Kalay, M. Sari :- {\it Int. J. Mod. Phys. A} {\bf 28}, 1350167 (2013).
\bibitem{stegr5} I. Mol :- {\it Advances in Applied Clifford Algebras} {\bf 27}, 2607 (2017).
\bibitem{dsit} T. Harko, T. S. Koivisto, F. S. N. Lobo, G. J. Olmo, D. Rubiera-Garcia :- {\it Phys. Rev. D} {\bf 98} 084043 (2018)
\bibitem{ftq1} Y. Xu, G. Li, T. Harko, S-D Liang :- {\it Eur. Phys. J. C.} {\bf 79}, 708 (2019)
\bibitem{mond1} O. Bertolami, C. G. Boehmer, T. Harko, F.S.N. Lobo :- {\it Phys. Rev. D} {\bf 75}, 104016 (2007)
\bibitem{stegr1} J. M. Nester, H.-J. Yo :- {\it Chin. J. Phys.} {\bf 37}, 113 (1999).
\bibitem{new1} J. Beltran Jimenez, L. Heisenberg, T. Koivisto :- {\it J. Cosmol. Astropart. Phys.} {\bf 08} 039 (2018).
\bibitem{dof1} Y. S. Myung :- {\it Adv.High Energy Phys.} {\bf 3901734} (2016)
\bibitem{dof2} M. Li, R-X. Miao, Y-G. Miao :- {\it JHEP} {\bf 1107} 108, (2011)
\bibitem{dof3} I. Soudi, G. Farrugia, V. Gakis, J. L. Said, E. N. Saridakis :- {\it Phys.Rev.D} {\bf 100} 4, 044008 (2019)
\bibitem{new2} H. Shabani, M. Farhoudi :- {\it Phys. Rev. D} {\bf 90} 044031 (2014)
\bibitem{new3} O. Bertolami, J. Paramos :- {\it J. Cosmol. Astropart. Phys.} {\bf 009} 1003 (2010)
\bibitem{new4} O. Bertolami, P. Frazao, J. Paramos :- {\it Phys. Rev. D} {\bf 86} 044034 (2012)
\bibitem{new5} M. Perucho :-{\it Galaxies} {\bf 7(3)} 70 (2019).
\bibitem{new6} T. Koivisto:- {\it Int. J. Geom. Methods Mod. Phys} {\bf 15}, No. supp 01, 1840006 (2018)
\bibitem{new7} H. F. M. Goenner :- {\it Found. Phys.} {\bf 14} 865 (1984)
\bibitem{new8} T. Koivisto :- {\it Class. Quant. Grav.} {\bf 23} 4289 (2006)
\bibitem{new9} O. Bertolami, C. G. Boehmer, T. Harko, F. S. N. Lobo :- {\it Phys. Rev. D} {\bf 75} 104016 (2007)
\bibitem{new10} M. Khodadi, A. Allahyari, S. Capozziello :- {\it Phys. Dark Univ.} {\bf 36} 101013 (2022)
\bibitem{new11} G. J. Olmo, D. Rubiera-Garcia :- {\it Phys. Lett. B} {\bf 740} 73 (2015)
\bibitem{lqg} C. Rovelli :- {\it Living Rev. Relativ.} {\bf 1}, 1 (1998)
\bibitem{brane} R. Maartens, K. Koyama :- {\it Living Rev. Relativity} {\bf 13}, 5 (2010)
\bibitem{dimless} T. Harko, T. S. Koivisto, F. S. N. Lobo, G. J. Olmo, D. R-Garcia :- {\it Phys. Rev. D} {\bf 98} 084043 (2018)
\bibitem{dimensions} P. G. L. Porta Mana :- {\it Eur. J. Phys.} {\bf 42} 045601 (2021), arxiv: 2007.14217
\bibitem{energyc1} S. Capozziello, S. Nojiri, S. D. Odintsov :- {\it Phys. Lett. B} {\bf 781}, 99 (2018).
\bibitem{hawk1} S. W. Hawking, G. F. R. Ellis :- The Large scale structure of space-time, {\it Cambridge University Press, Cambridge} (1973)
\bibitem{nec1} S. Capozziello, F. S. N. Lobo, J. P. Mimoso :- {\it Phys. Rev. D} {\bf 91} 124019 (2015)
\bibitem{nec2} S. Capozziello, F. S. N. Lobo, J. P. Mimoso :- {\it Phys. Lett. B} {\bf 730} 280 (2014)
\bibitem{nec3} S. Mandal, P. K. Sahoo, J. R. L. Santos :- {\it Phys. Rev. D} {\bf 102} 024057 (2020)
\bibitem{nec4} S. Arora, J. R. L. Santos, P. K. Sahoo :- {\it Phys. Dark Univ.} {\bf 31} 100790 (2021)
\bibitem{nec5} K. Bamba, M. Ilyas, M. Z. Bhatti, Z. Yousaf :- {\it Gen. Rel. Grav.} {\bf 49} 8 (2017)
\bibitem{nec6} J. Santos, J. S. Alcaniz, M. J. Reboucas F. C. Carvalho :- {\it Phys. Rev. D} {\bf 76}, 083513 (2007)
\bibitem{nec7} N. M. Garcia, T. Harko, F. S. N. Lobo, J. P. Mimoso :- {\it Phys. Rev. D} {\bf 83}, 104032 (2011)
\bibitem{planck} Planck Collaboration :- {\it Astron. Astrophys.} {\bf 641}, A6 (2020)
\bibitem{capo2} S. Capozziello, R. D'Agostino, O. Luongo :- {\it Int. J. Mod. Phys. D} {\bf 28}, 1930016 (2019).
\bibitem{nasa} WMAP's Universe : National Aeronautics and Space Administration website. (Link:
$https://wmap.gsfc.nasa.gov/universe/uni_matter.html$)
\bibitem{ec1} M. Visser, C. Barcelo :- {\it Cosmo-99} {\bf 98} (2000), [arxiv: 0001099].
\bibitem{ec2} E. E. Flanagan, R. M. Wald :- {\it Phys. Rev. D} {\bf 54}, 6233 (1996)
\bibitem{ec3} C. Barcelo, M. Visser :- {\it Phys. Lett. B} {\bf 466}, 127 (1999)
\bibitem{ep1} E. Di Casola, S. Liberati, S. Sonego :- {\it Am. J. Phys.} {\bf 83}, 39 (2015)
\bibitem{ep2} C. M. Will :- Theory and Experiment in Gravitational Physics, Second Edition {\it (Cambridge, England: Cambridge University Press)} (2018)
\bibitem{ep3} J. Magueijo, L. Smolin :- {\it Class. Quantum Grav.} {\bf 21} 1725 (2004)
\bibitem{ep4} J. Magueijo, L. Smolin :- {\it Phys. Rev. Lett.} {\bf 88} 19 (2002)
\bibitem{ep5} G. Voisin et. al. :- {\it Astronomy \& Astrophysics} {\bf 638} A24 (2020)
\bibitem{ep6} R. D. Reasenberg, J. D. Phillips :- {\it Int. J. Mod. Phys. D} {\bf 16}, 2245 (2007)
\bibitem{ep7} O. Bertolami, F. Gil Pedro, M. Le Delliou :- {\it Phys. Lett. B} {\bf 654}, 165 (2007)


\end{thebibliography}
\end{document}